\documentclass[11pt]{article}
\usepackage{latexsym}
\usepackage{amsmath}
\usepackage{amsthm}
\usepackage{amsfonts}
\usepackage{amsxtra}
\usepackage[lite]{amsrefs}
\usepackage{hyperref}
\allowdisplaybreaks
\title{Many-body Hamiltonians in implicitly defined frames}
\author{Antonio O.\ Bouzas\thanks{abouzas@mda.cinvestav.mx}
  \\\small
  Departamento de F\'{\i}sica Aplicada, CINVESTAV-IPN
  \\[-0.5ex]\small Carretera Antigua a Progreso Km.\ 6,
  Apdo.\ Postal 73 ``Cordemex''\\[-0.5ex]\small M\'erida 97310,
  Yucat\'an, M\'exico}
\date{\today}
\newcommand{\Ll}{\ensuremath{\mathcal{L}}}

\renewcommand{\H}{\ensuremath{\mathcal{H}}}
\newcommand{\K}{\ensuremath{\mathcal{K}}}
\newcommand{\Kz}{\ensuremath{\mathcal{K}_0}}

\newcommand{\Ko}{\ensuremath{\mathcal{K}_{1}}}
\newcommand{\Kt}{\ensuremath{\mathcal{K}_{2}}}
\newcommand{\mat}[1]{\ensuremath{\boldsymbol{#1}}}
\newcommand{\matd}[1]{\ensuremath{\boldsymbol{\dot{#1}}}}
\newcommand{\ver}[1]{\ensuremath{\boldsymbol{\widehat{#1}}}}

\newcommand{\Q}{\ensuremath{\mathcal{Q}}}
\newcommand{\R}{\ensuremath{\mathcal{R}}}
\renewcommand{\P}{\ensuremath{\mathcal{P}}}
\newcommand{\C}{\ensuremath{\mathcal{C}}}
\newcommand{\G}{\ensuremath{\mathcal{G}}}
\newcommand{\D}{\ensuremath{\mathcal{D}}}
\newcommand{\B}{\ensuremath{\mathcal{B}}}
\newcommand{\J}{\ensuremath{\mathcal{J}}}
\newcommand{\N}{\ensuremath{\mathcal{N}}}
\newcommand{\F}{\ensuremath{\mathcal{F}}}
\newcommand{\vrcm}{\ensuremath{\mat{r}_\mathrm{cm}}}
\newcommand{\vdrcm}{\ensuremath{\matd{r}_\mathrm{cm}}}
\newcommand{\rcm}{\ensuremath{r_\mathrm{cm}}}
\newcommand{\vlcm}{\ensuremath{\mat{l}_\mathrm{cm}}}
\newcommand{\lcm}{\ensuremath{l_\mathrm{cm}}}
\newcommand{\pth}[1]{p_{\theta_{#1}}}
\begin{document}
\maketitle
\begin{abstract}
  We study the quantization of three-dimensional many-body systems in
  rotating coordinate frames defined implicitly by frame conditions.
  We carry out the elimination of orientational degrees of freedom in
  general, giving the Hamiltonian for the $N$-particle system in a
  broad class of body frames in terms of frame conditions and internal
  coordinates.  We obtain several forms for the kinetic energy
  operator and compare them to related expressions in the literature.
\end{abstract}
\section{Introduction}
\label{sec:intro}

The problem of separating the dynamics of quantum many-body systems
into collective rotations and internal motions leads to their
quantization in rotating frames.  We refer to a rotating frame as a
body frame when the components of the total angular momentum operator
in a space-fixed frame and in the rotating frame satisfy the same
commutator algebra as in the rigid body problem.  In the latter case
the body frame is essentially unique, up to time-independent rotations
and symmetry transformations of the rigid body.  For a general
$N$-particle system there is a large freedom to choose a body frame.
It is thus of interest to study the quantization of many body systems
in a class of body frames as wide as possible.

In this paper we study the quantization of three-dimensional many-body
systems in rotating coordinate frames defined in implicit form by
frame, or gauge, conditions.  We carry out the elimination of
orientational degrees of freedom in general, giving the Hamiltonian
for the $N$-particle system in a broad class of body frames in terms
of frame conditions and internal coordinates.  We obtain several forms
for the kinetic energy operator and compare them to related
expressions in the literature, showing how the coefficient functions
are fixed by the frame convention, through frame conditions, and
internal coordinates.  In the case of linear frames and body-frame
coordinates our results reduce to those previously obtained in
\cite{bou04}.

The generic Hamiltonians discussed here can be applied to specific
physical systems by choosing internal coordinates appropriate to the
system under consideration.  Those physical problems include, for
instance, the determination of molecular rotation-vibration energy
levels and their wave functions \cite{han87}, and scattering problems
in molecular, atomic and nuclear physics. There is a vast body of
literature on the quantization of many-body systems which we do not
try to summarize here.  Broad reviews relevant to the point of view
adopted in this paper are given in \cites{lit97,mer03}.

The outline of the paper is the following.  In the next section we
discuss several technical issues related to body frames and the frame
conditions defining them that are needed in order to obtain the
Hamiltonian and quantum inner product for a many body system.  Those
include the form of admissible frame conditions and their
reparameterizations, the body-frame angular momentum, and internal
coordinates.  In section \ref{sec:stand} we derive the body-frame
kinetic energy operator in standard order in terms of internal
coordinates and frame conditions.  The form of wave functions referred
to the body frame and their inner product is discussed in section
\ref{sec:matrig}, where we also give the representations of the total
angular momentum and kinetic energy operators both as irreducible
matrices and in rigid rotator form.  In section \ref{sec:other} we
give two alternate forms for the Hamiltonian and discuss their
equivalence with the standard ordered form given in section
\ref{sec:stand}.  Also, we make contact with the gauge-field formalism
of \cite{lit97} by locally expressing gauge fields in terms of frame
conditions.  Examples with $N=3$ and 4 are briefly examined in
sections \ref{sec:examp} and \ref{sec:examp4} as verifications of the
formalism of the previous sections, and the results compared with
those from the molecular literature.  In section \ref{sec:final} we
give some final remarks.

\section{Preliminaries}
\label{sec:preli}

We consider rotating frames whose definition depends only on the
coordinates of the particles and not on their velocities, nor on the
angular velocity of the frame itself.  Those frames can always be
defined implicitly by imposing conditions on the position vectors of
the particles.  Most of the body frames commonly used in the
literature belong to this class, as illustrated in sections
\ref{sec:examp} and \ref{sec:examp4} below with two familiar examples.
Another well-known example is the Eckart frame \cites{eck35,lou76}.
We do not impose any restriction on the form of frame conditions,
provided they fix the frame uniquely.  The particular case of frame
conditions depending linearly on the particles coordinates was
considered in \cite{bou04} from the point of view of gauge invariance.

\subsection{Frame conditions and their reparameterizations}
\label{sec:frame}

The frame conditions defining the body frame must fix its six
degrees of freedom.  The translational degrees of freedom are fixed by
choosing the center of mass frame.  Thus, the frame conditions take
the form,\footnote{The letters $a,b,c,d$ are used for non-tensorial
  indices, as in $\G_a$.  Summation over those indices and their
  ranges of variation are always explicitly indicated.  We only use
  the summation convention for tensor indices, denoted by latin
  letters $i,j,k,l,\ldots$, which always run from 1 to 3.  Greek
  indices number particles.}
\begin{equation}
  \label{eq:frame}
\mat{\C}(\{\mat{r}\}) \equiv \sum_{\alpha=1}^N m_\alpha
\mat{r}_\alpha=0~,
\qquad
\G_a(\{\mat{r}\}) = 0~,
\quad
a=1,2,3~,
\end{equation}
where $\G_a$ are three conditions fixing the orientational degrees of
freedom.  We denote by $\{\mat{r}\} = \{\mat{r}_1, \ldots,
\mat{r}_N\}$ a generic configuration\footnote{A configuration of the
  system is actually $\{\mat{r},\matd{r}\}$.  We refer to
  $\{\mat{r}\}$ as a configuration here for convenience.} of the
$N$-particle system.  Since lab-frame configurations are not
restricted, any $\{\mat{r}\}$ can be a lab-frame configuration.  Those
configurations satisfying the frame conditions are denoted by
$\{\mat{R}\}$ .  Thus, $\mat{\C}(\{\mat{R}\}) \equiv 0 \equiv
\G_a(\{\mat{R}\})$ and such a set $\{\mat{R}\}$ of $N$ position
vectors can be a body-frame configuration.  We introduce also the
following notations,
\begin{equation}
  \label{eq:frnot}
  \frac{\partial \G_a}{\partial \mat{R}_\alpha} \equiv \frac{\partial
    \G_a}{\partial \mat{r}_\alpha}(\{\mat{R}\})~,
  \qquad 
  \Q_{ai}(\{\mat{R}\}) \equiv \sum_{\alpha=1}^N \frac{\partial
    \G_a}{\partial R_{\alpha j}} \varepsilon_{jik} R_{\alpha k}~, 
  \qquad
  \R^2_{ab}(\{\mat{R}\}) \equiv \sum_{\alpha=1}^N \frac{1}{m_\alpha}
  \frac{\partial \G_a}{\partial R_{\alpha l}} \frac{\partial
    \G_b}{\partial R_{\alpha l}} ~
\end{equation}
which will be used throughout the paper.

In order for the conditions $\G_a=0$ to fix the orientational freedom
they must not be rotation invariant.  Thus, they must satisfy the
admissibility condition,
\begin{equation}
  \label{eq:admi}
  \det(\Q_{ai}(\{\mat{R}\})) \neq 0
  \qquad \text{for}\qquad
  \G_a(\{\mat{R}\}) = 0~,
\end{equation}
except maybe at singular configurations.  We also assume that, except
for singular configurations, the relation $
\det(\R_{ab}^2(\{\mat{R}\}))$ $\neq 0$ for $\G_a(\{\mat{R}\}) = 0$ is
fulfilled so that the frame manifold possesses a tangent space at
$\{\mat{R}\}$.  Typically, $\{\mat{R}_\alpha = 0\}$ is a singular
configuration.  Furthermore, for the two sets of conditions in
(\ref{eq:frame}) to be compatible the rotational conditions $\G_a$
must be translation invariant, $\G_a(\{\mat{r}_\alpha + \mat{v}\}) =
\G_a(\{\mat{r}_\alpha\})$ for any \mat{v}.  This is satisfied by all
usual frame conditions (see, e.g., \cites{bou04,eck35}).  In what
follows, however, it will be enough to assume only the weaker form,
\begin{equation}
  \label{eq:weak}
  \sum_{\alpha=1}^N \frac{\partial\G_a}{\partial R_{\alpha j}}=0~. 
\end{equation}
The condition $\mat{\C}=0$, on the other hand, is clearly rotation
invariant.  

The frame conditions are obviously not unique.  Consider the class of
reparameterizations,
\begin{equation}
  \label{eq:reparam}
  \G_a'(\{\mat{r}\}) = \sum_{b=1}^3 \P_{ab}(\{\mat{r}\})
  \G_b(\{\mat{r}\})~, 
\end{equation}
where $\P_{ab}(\{\mat{r}\})$ is non-singular on the frame
manifold,
\begin{equation}
  \label{eq:nsing}
  \det(\P_{ab}(\{\mat{R}\})) \neq 0 \qquad \text{for} \qquad
  \G_c(\{\mat{R}\}) = 0~.
\end{equation}
The frame conditions $\G_a'$ define the same frame as $\G_a$ and we
have,
\begin{equation}
  \label{eq:reparam2}
  \R^{\prime 2}_{cd} = \sum_{\alpha=1}^N \frac{1}{m_\alpha}
  \frac{\partial\G'_c}{R_{\alpha j}} \frac{\partial\G'_d}{R_{\alpha
      j}} = \sum_{c_1, d_1=0}^3 \P_{cc_1} \P_{dd_1} \R^2_{c_1d_1}~,
  \qquad
  \Q^\prime_{ai} = \sum_{b=1}^3 \P_{ab} \Q_{bi}~.
\end{equation}
Notice that the reparameterization (\ref{eq:reparam}) is not
necessarily linear in the frame conditions, since the coefficients
$\P_{ab}$ can depend on $\G_a$.  For instance $\P_{ab}=\R^{-2}_{ab}$
satisfies (\ref{eq:nsing}) by assumption, and leads to $\G'_a$
orthonormal on the frame manifold, \emph{i.e.}, $\R^{\prime 2}_{ab} =
\delta_{ab}$. Thus, we could assume without loss of generality that
frame conditions are orthonormal in this sense.  We shall not do so,
however, because orthonormalizing a set of frame conditions can be
inconvenient in practice.

Another important example is furnished by the reparameterization
\begin{equation}
  \label{eq:spreparam}
  \F_i(\{\mat{r}\}) = \sum_{b=1}^3
  \Q^{-1}_{ib}(\{\mat{r}\}) \G_b(\{\mat{r}\})~. 
\end{equation}
From (\ref{eq:admi}) we see that (\ref{eq:nsing}) is satisfied.
Clearly, $\F_i$ depend non-linearly on the $\G_a$. If $\G_a'$ are any
frame conditions equivalent to the $\G_a$ then, in a neighborhood of
the frame manifold $\G_a = 0$, they can be related by a
reparameterization of the form (\ref{eq:reparam}).  Using
(\ref{eq:reparam2}) and (\ref{eq:spreparam}) we see that the frame
conditions $\F_i$ are invariant under reparameterizations of $\G_a$,   
\begin{equation}
  \label{eq:equiv}
  \F'_i(\{\mat{r}\}) \equiv \sum_{b=1}^3
  \Q^{\prime -1}_{ib}(\{\mat{r}\}) \G_b'(\{\mat{r}\}) =
  \sum_{c=1}^3 \Q^{-1}_{ic}(\{\mat{r}\})
  \G_c(\{\mat{r}\}) = \F_i(\{\mat{r}\})~.
\end{equation}
Other reparameterization-invariant quantities involving the frame
conditions $\G_a$ can be expressed most economically in terms of
$\F_i$ and their derivatives
\begin{gather}
  \F_{i\alpha j}\equiv \frac{\partial\F_i}{\partial R_{\alpha j}} =
  \sum_{a=1}^3 \Q^{-1}_{ia} \frac{\partial\G_a}{\partial R_{\alpha
      j}}    \label{eq:sprep2}~.
\end{gather}
These quantities play an important role in what follows since, as
shown below, the frame conditions $\G_a$ enter the Hamiltonian only
through $\F_i$ and their derivatives.  This ensures that \H\ is
reparameterization invariant, as all observables should be.  

The previous analysis can be extended to a more general class of
reparameterizations of the frame conditions $\G_a' = \P_a[\{\G_b\}]$,
involving arbitray functionals $\P_a$ of $\G_b$ which are not singular
at $\G_b=0$.  Clearly, only the behavior of the new frame conditions
$\G_a'$ in a small neighborhood of the frame manifold $\G_a=0$ is
relevant.  In such neighborhood we can expand $\P_a$ about $\G_a=0$,
thus obtaining a reparameterization of the form (\ref{eq:reparam}).

\subsection{Body frame transformation and angular momentum}

The transformation relating a configuration $\{\mat{r}\}$ of
the system in the laboratory frame to the corresponding configuration
$\{\mat{R}\}$ in the body frame defined by the conditions
(\ref{eq:frame}) is 
\begin{equation}
  \label{eq:trans}
  \mat{R}_\alpha = \mat{U}(\mat{r}_\alpha - \vrcm)
\end{equation}
with \vrcm\ the center-of-mass position in the lab frame and
$\mat{U}=\mat{U}(\{\theta_a\})$ an orthogonal matrix parameterized by
three angular variables $\{\theta_a\}_{a=1}^3$.  Although our approach
and results do not depend on any specific parameterization of the
rotation group, some parameterization-dependent quantities such as the
momenta $p_{\theta_a}$ conjugate to $\theta_a$ are physically
meaningful and play an important role in some intermediate
calculations.  All the information we will need about the
parameterization of \mat{U}\ is encoded in the matrices \mat{\Lambda}\ 
and \mat{\lambda}\ defined by
\begin{equation}
  \label{eq:lambdas}
  \frac{\partial\mat{U}}{\partial \theta_a} \mat{U}^\dagger =
  \Lambda_{ai} \mat{T}_i~,
  \qquad
  \mat{U}^\dagger \frac{\partial\mat{U}}{\partial \theta_a} =
  \lambda_{ai} \mat{T}_i~,
  \qquad
  a=1,2,3,
\end{equation}
where $\mat{U}^\dagger$ is the tranpose of $\mat{U}$ and $\mat{T}_j$
are the standard generators of the $so(3)$ algebra,
$(\mat{T}_j)_{ik}=\varepsilon_{ijk}$.  The three matrices
$\partial\mat{U}/\partial\theta_a \mat{U}^\dagger$, $a=1,2,3$, must be
a basis of $so(3)$ for all values of $\{\theta_b\}$ if the
parameterization is to be well defined.  Thus, the matrix
$\Lambda_{ai}$ is invertible and, analogously, so is $\lambda_{ai}$.
{F}rom the unimodularity of \mat{U}\ it follows that $\mat{U}^\dagger
\mat{T}_i \mat{U} = U_{ij} \mat{T}_j$ and then, from
(\ref{eq:lambdas}), $\lambda_{aj} = \Lambda_{ai} U_{ij}$.

The frame conditions (\ref{eq:frame}) determine the time dependence of
$\{\theta_a\}$ so that, given a trajectory $\{\mat{r}_\alpha(t)\}$ of
the system in the lab frame, we have $\G_a(\{\mat{R}_\alpha(t)\}) =
\G_a(\{\mat{U}(\{\theta_a(t)\})(\mat{r}_\alpha(t)-\vrcm)\}) = 0$ for
all $t$.  {F}rom (\ref{eq:trans}) we then have, with $M$ the total
mass of the system,
\begin{equation}
  \label{eq:dR/dr}
  \frac{\partial R_{\beta i}}{\partial r_{\alpha j}} =
  \left(\delta_{\alpha\beta} - \frac{m_\alpha}{M}\right) U_{ij} +
  \frac{\partial U_{ik}}{\partial r_{\alpha j}} U_{lk} R_{\beta l}~. 
\end{equation}
Substituting (\ref{eq:dR/dr}) into the relation $\partial
\G_a/\partial R_{\alpha j} = 0$, and using the definition
(\ref{eq:frnot}) for \Q\ and the antisymmetry of $(\partial
U_{ik}/\partial r_{\alpha j} U_{lk})$ in $i$ and $l$, we obtain the
relation
\begin{equation}
  \label{eq:Ufund}
  \frac{\partial U_{ik}}{\partial r_{\alpha j}} U_{lk} = \sum_{a=1}^3 
  \varepsilon_{ilm} \Q_{ma}^{-1} \frac{\partial\G_a}{\partial
  R_{\alpha n}}U_{nj} = \varepsilon_{ilm}\F_{m\alpha n} U_{nj}~, 
\end{equation}
which expresses $\partial\mat{U}/\partial r_{\alpha j}
\mat{U}^\dagger$ in terms of $\{\mat{R}_\gamma\}$ and $\{\theta_b\}$.
This expression characterizes the dependence of \mat{U} on
$\{\mat{r}_\alpha\}$, and will be important below, especially in the
discussion of angular momentum.  Some further consequences of
(\ref{eq:Ufund}) are discussed in appendix \ref{sec:app0}.  Together,
(\ref{eq:dR/dr}) and (\ref{eq:Ufund}) lead to
\begin{equation}
  \label{eq:dR/dr2}
  \frac{\partial R_{\beta i}}{\partial r_{\alpha j}} =
  \left(\delta_{\alpha\beta} - \frac{m_\alpha}{M}\right) U_{ij} +
  \varepsilon_{ikl}\F_{l\alpha m} U_{mj} R_{\beta k}~, 
\end{equation}
which will also be useful below.

{F}rom (\ref{eq:trans}) and $\partial R_{\alpha i}/\partial \theta_a
=0$ we get
\begin{equation}
  \label{eq:dr/dtheta}
  \frac{\partial r_{\alpha i}}{\partial\theta_a} =
  \frac{\partial U_{ki}}{\partial \theta_a} U_{kj} (r_{\alpha j} -
  {\rcm}_j)~. 
\end{equation}
We assume that interactions among particles do not depend on their
velocities.  Classically, the momenta $p_{\theta_a}$ conjugate to
$\theta_a$ is then
$p_{\theta_a}=\partial\Ll/\partial\dot{\theta}_a =
\sum_{\alpha=1}^N (\partial r_{\alpha i}/\partial\theta_a) 
(\partial\Ll/\partial\dot{r}_{\alpha i})$
where \Ll\ is the classical Lagrangian in the lab frame.  Thus, taking
(\ref{eq:lambdas}) and (\ref{eq:dr/dtheta}) into account, we have
\begin{equation}\label{eq:angmom}
p_{\theta_a} = -\lambda_{ai} (l_i - {\lcm}_i) = -\Lambda_{aj} L_j
\qquad\mathrm{with}\quad
\vlcm \equiv M \vrcm\wedge\vdrcm~,
\quad
\mat{L} = \mat{U} (\mat{l} - \vlcm)~. 
\end{equation}
\vlcm\ is the center-of-mass angular momentum in the lab frame, and
\mat{L}\ the total angular momentum about the center of mass in the
moving frame.

In the lab frame the angular momentum operator \mat{l}\ satisfies the
usual commutator algebra.  Using (\ref{eq:Ufund}), the definition
(\ref{eq:frnot}) of \Q, and the unimodularity of \mat{U}, we obtain,
\begin{equation}
  \label{eq:commlU}
  [l_i,U_{jk}] = \sum_{\alpha=1}^N \varepsilon_{ilm} r_{\alpha
  l}\frac{1}{i} \frac{\partial U_{jk}}{\partial r_{\alpha m}} = i
  \varepsilon_{ikn} U_{jn}~. 
\end{equation}
Using (\ref{eq:commlU}) and the definition (\ref{eq:angmom}) of
\mat{L}, its commutators can now be computed
\begin{equation}
  \label{eq:commute}
  [L_i,U_{jk}] = -i\varepsilon_{ijn} U_{nk}~,
  \quad
  [l_i, L_j]=0~,
  \quad
  [L_i, L_j] = -i\varepsilon_{ijk} L_k~.
\end{equation}
The commutators among components of \mat{l}\ and \mat{L}\ are the same
as those for a rigid body, with \mat{l}\ the angular momentum in the
laboratory and \mat{L}\ in the body frame.   We notice also that
$[l_i, R_{\alpha j}]=0=[L_i, R_{\alpha j}]$ as expected, since in the
body frame rotations act only on the angles $\{\theta_a\}$. 

\subsection{Internal coordinates}
\label{sec:intcoord}

In order to describe the dynamics we introduce a set of $3N-6$
internal coordinates $\{t_a\}_{a=1}^{3N-6}$ defined locally as
independent rotation- and translation-invariant functions of
configuration space $t_a = t_a(\{\mat{r}\})$.  Some consequences of
the Euclidean invariance of $t_a$ which will be used below are
$t_a(\{\mat{r}\}) = t_a(\{\mat{R}\})$ and, introducing notations
analogous to the first of (\ref{eq:frnot}),
\begin{equation}
  \label{eq:euclid}
  \frac{\partial t_a}{\partial\mat{R}_\alpha}\equiv \frac{\partial
    t_a}{\partial\mat{r}_\alpha}(\{\mat{R}\}) = \mat{U}
  \frac{\partial
    t_a}{\partial\mat{r}_\alpha}(\{\mat{r}\})~,\\
  \quad
  \frac{\partial^2 t_a}{\partial R_{\alpha j}\partial R_{\alpha j}}
  \equiv \frac{\partial^2 t_a}{\partial r_{\alpha j}\partial r_{\alpha 
    j}}(\{\mat{R}\}) = \frac{\partial^2 t_a}{\partial r_{\alpha
    j}\partial r_{\alpha j}}(\{\mat{r}\})
\end{equation}
which can be derived using (\ref{eq:trans}) and (\ref{eq:Ufund}).  The
body frame, specified by the frame conditions (\ref{eq:frame}), fixes
relations of the form $\mat{R}_\alpha = \mat{R}_\alpha(\{t_a\})$ such
that the conditions (\ref{eq:frame}) are satisfied identically when
evaluated on $\mat{R}_\alpha(\{t_a\})$.  Thus, the functions
$\mat{R}_\alpha(\{t_a\})$ are a parametric solution to the conditions
(\ref{eq:frame}).  Through the relation inverse to (\ref{eq:trans}),
$\mat{r}_\alpha = \mat{U}^\dagger(\{\theta_a\})
\mat{R}_\alpha(\{t_b\})+\vrcm$, the internal coordinates $\{t_b\}$
together with the orientational and translational ones, $\{\theta_a\}$
and $\vrcm$ resp., give a set of $3N$ local coordinates in
configurations space.

Usually, internal coordinates are given as $3N-6$ independent
functions $\{t_a\}$ of as many independent Euclidean invariants chosen
out of the set of all dot and triple products among $\{\mat{r}_\alpha
- \mat{r}_\beta\}_{\alpha,\beta=1}^N$.  One way of introducing local
coordinates in configuration space is to start with a parametric
solution $\{\mat{R}(\{Q\})\}$ to the frame conditions, where
$\{Q\}=\{Q_a\}_{a=1}^{3N-6}$ is a set of independent parameters.
These parametric relations can be inverted to give $3N-6$ local
coordinates $Q_a(\{\mat{R}\})$ on the body frame.  Such ``inverse'' is
clearly not unique, however, since at each point $\{\mat{R}\}$ in the
body frame the functions
\begin{equation*}
Q'_a(\{\mat{r}\}) = Q_a(\{\mat{r}\})
+ \sum_{b=1}^3 \lambda_{ab}(\{\mat{r}\}) \G_b(\{\mat{r}\}) +
\mat{\lambda}_a(\{\mat{r}\})\cdot\C(\{\mat{r}\}) 
\end{equation*}
with $\lambda_{ab}$ and $\mat{\lambda}_a$ arbitrary coefficient
functions, take on the same values as $Q_a(\{\mat{r}\})$.  That
ambiguity can be fixed by imposing additional conditions such as
$\sum_{\alpha=1}^3\partial Q_a/\partial\mat{R}_\alpha = 0$ and either
\begin{subequations}
  \begin{align}
    \label{eq:add1}
    \sum_{\alpha=1}^3\frac{1}{m_\alpha}
    \frac{\partial \G_a}{\partial\mat{R}_\alpha}
    \frac{\partial Q_b}{\partial\mat{R}_\alpha} &= 0~,
    \qquad 1\leq a \leq 3~, \quad 1\leq b\leq 3N-6\\
    \intertext{or}
    \sum_{\alpha=1}^3 \mat{R}_\alpha\wedge
    \frac{\partial Q_b}{\partial\mat{R}_\alpha} &= 0~,
    \qquad 1\leq b\leq 3N-6~.
  \end{align}
\end{subequations}
The first set of conditions fixes the ambiguity because
$\det(\R^2(\{\mat{R}\})) \neq 0$, and the second because
$\det(\Q_{ai}(\{\mat{R}\})) \neq 0$, by assumption.  Once the
ambiguity has been fixed, to each configuration $\{\mat{R}\}$ in the
body frame there corresponds one and (except at singular points) only
one set of parameters $Q_a(\{\mat{R}\})$.  We call those parameters
body-frame coordinates.  Rotation- and translation-invariant
coordinates in configuration space can be obtained locally by
extending the body-frame coordinates by rotation and translation,
$t_a(\{\mat{r}\}) = Q_a(\{\mat{R}(\{\mat{r}\})\})$, with
$\mat{R}_\alpha(\{\mat{r}\})$ given by (\ref{eq:trans}).  These $t_a$
are invariant under Euclidean motions because
$\{\mat{R}(\{\mat{r}\})\}$~are.

The body frame coordinates $Q_a$, considered as local functions of
configuration space $Q_a(\{\mat{r}\})$ are not, in general, rotation
or translation invariant.  For instance, when the frame conditions are
linear the $Q_a(\{\mat{r}\})$ can be chosen to be linear functions of
$\{\mat{r}\}$, therefore not rotation invariant, as in the case of
normal coordinates in the Eckart frame.  A simple example is given in
section \ref{sec:linfram}. The derivation of the Hamiltonian operator
and Hilbert-space inner product in terms of linear body-frame
coordinates and their conjugate momenta has been discussed in detail
in \cites{bou04,men03}.  In this paper we confine ourselves to a
dynamical description in terms of Euclidean-invariant internal
coordinates $\{t_a\}$.

\subsection{Hamiltonian operator in configuration space}
\label{sec:hamconf}

The kinetic-energy operator in the lab frame is given by the familiar
expression 
\begin{equation}
  \label{eq:fam}
  \K = -\frac{1}{2} \sum_{\alpha=1}^N \frac{1}{m_\alpha}
  \frac{\partial^2}{\partial r_{\alpha j}\partial r_{\alpha j}}~.
\end{equation}
In this equation and in what follows we use units such that $\hbar=1$
unless otherwise stated.  In terms of mass-weighted position vectors
$\mat{r}'_\alpha = \sqrt{m_\alpha}\mat{r}_\alpha$, \K\ takes the form
of ($-1/2$ times) a Laplacian operator in Cartesian coordinates in
$3N$ dimensional space.  Given a set $\{q\}=\{q_1,\ldots,q_{3N}\}$ of
curvilinear coordinates in configuration space we can write \K\ as a
Laplacian in either one of two commonly used forms.  Application of
the chain rule to (\ref{eq:fam}) leads to the standard ordering, with
all derivative operators to the right of coefficient functions,
\begin{equation}
  \label{eq:standard}  
\begin{gathered}
  \K = -\frac{1}{2}\sum_{a,b=1}^{3N} k^{-1}_{ab}
  \frac{\partial}{\partial q_a}\frac{\partial}{\partial q_b} -
  \frac{1}{2}\sum_{b=1}^{3N} k_b
  \frac{\partial}{\partial q_b}~,\\
  k^{-1}_{ab} = \sum_{\alpha=1}^N \frac{1}{m_\alpha} \frac{\partial
    q_a}{\partial r_{\alpha j}} \frac{\partial q_b}{\partial r_{\alpha
      j}}~,
  \quad
  k_{ab} = \sum_{\alpha=1}^N m_\alpha \frac{\partial
    r_{\alpha j}}{\partial q_a} \frac{\partial r_{\alpha j}}{\partial
    q_b}~,
  \quad
  k_b = \sum_{\alpha=1}^N \frac{1}{m_\alpha}\frac{\partial^2
    q_b}{\partial r_{\alpha j}\partial r_{\alpha j}}~.
\end{gathered}
\end{equation}
This form for the kinetic energy has been widely used in molecular
physics and leads to expressions which are usually simpler than in
Weyl ordering.  An equivalent expression is
\begin{equation}
  \label{eq:covariant}
  \K = -\frac{1}{2J} \sum_{a,b=1}^{3N} \frac{\partial}{\partial q_a}
  k^{-1}_{ab} J \frac{\partial}{\partial q_b}~,
  \qquad
  J = \det(k_{ab})^{1/2}~.
\end{equation}
Despite their equivalence, (\ref{eq:standard}) and
(\ref{eq:covariant}) lead to considerably different forms for
the kinetic energy operator for many-body systems, especially after
the momenta conjugate to orientational variables are eliminated in
favor of the total angular momentum.  Those forms, and the
relations among them, are discussed below in sections \ref{sec:stand}
and \ref{sec:other}.

\section{Hamiltonian operator in standard ordering}
\label{sec:stand}

The kinetic energy can be expressed in terms of internal coordinates
within a given frame convention (\ref{eq:frame}) by taking $\{q\}$ in
(\ref{eq:standard}) to be the union of the set $\{t_a\}_{a=1}^{3N-6}$
with the orientational coordinates $\{\theta_b\}_{b=1}^3$ and the lab
frame center-of-mass vector \vrcm.  The kinetic energy term depending
on \vrcm\ trivially decouples from the other degrees of freedom so we
simply ignore it in what follows.  We introduce the notations $p_a =
-i\partial/\partial t_a$ ($a=1,\ldots,3N-6$),
$\pth{b}=-i\partial/\partial\theta_b$ ($b=1,2,3$) and
$\K=\Kz+\K_{\theta1} +\K_{\theta2}$, where \Kz, the vibrational
kinetic energy, does not contain $\pth{a}$, $\K_{\theta1}$ contains
the term quadratic in $\pth{a}$ and those terms linear in $\pth{a}$
whose coefficients involve only first derivatives of $\theta_a$ with
respect to $\{\mat{r}\}$, and $\K_{\theta2}$ gathers the remaining
terms linear in $\pth{a}$, with coefficients given by second
derivatives of $\theta_a$.

An expression for \Kz\ can be immediately obtained from
(\ref{eq:standard}) as,
\begin{equation}
  \label{eq:kostan}  
\begin{gathered}
  \Kz = \frac{1}{2}\sum_{a,b=1}^{3N-6} g_{ab} p_a p_b +
  \frac{1}{2i}\sum_{b=1}^{3N-6} g_b p_b~,\\
  g_{ab} = \sum_{\alpha=1}^N \frac{1}{m_\alpha} \frac{\partial
  t_a}{\partial R_{\alpha j}} \frac{\partial t_b}{\partial R_{\alpha 
      j}}~,
  \quad
  g_b = \sum_{\alpha=1}^N \frac{1}{m_\alpha}\frac{\partial^2
    t_b}{\partial R_{\alpha j}\partial R_{\alpha j}}~,
\end{gathered}
\end{equation}
where use was made of (\ref{eq:euclid}).  Notice that $g_{ab}$ and
$g_b$ depend only on $\{t_a\}$.

In order to obtain $\K_{\theta1}$ we need to specify the dependence of 
the orientational coordinates $\{\theta_a\}$ on the lab frame
coordinates $\{\mat{r}\}$.  Using (\ref{eq:lambdas}) we have
\begin{equation}
  \label{eq:lambdas2}
  \frac{\partial U_{ik}}{\partial r_{\alpha j}} U_{lk} =
  \sum_{b=1}^3 \frac{\partial \theta_b}{\partial r_{\alpha j}}
  \Lambda_{bm} \varepsilon_{iml}~.
\end{equation}
Equating the r.h.s.\ of (\ref{eq:lambdas2}) to that of
(\ref{eq:Ufund}), we obtain the dependence of $\theta_a$ on lab frame
coordinates in terms of the frame conditions (\ref{eq:frame}),
\begin{equation}
  \label{eq:thetafund}
  \frac{\partial\theta_a}{\partial r_{\alpha j}} = -\Lambda^{-1}_{ma}
  \F_{m\alpha i} U_{ij}~,
  \qquad a=1,2,3~.
\end{equation}
This expression is independent of the frame-conditions
parameterization, as it should be.  From (\ref{eq:thetafund}) and
(\ref{eq:euclid}), the corresponding blocks in the matrix
$k^{-1}_{ab}$ defined in (\ref{eq:standard}) are
\begin{equation}
  \label{eq:blocks1}
  \begin{gathered}
    k^{-1}_{a\theta_b} = \sum_{\alpha=1}^N \frac{1}{m_\alpha}
    \frac{\partial t_a}{\partial r_{\alpha j}} \frac{\partial
      \theta_b}{\partial r_{\alpha j}} = -\sum_{\alpha=1}^N
    \frac{1}{m_\alpha} \F_{m\alpha j} \frac{\partial
      t_a}{\partial R_{\alpha j}} \Lambda^{-1}_{mb}\\
    k^{-1}_{\theta_a\theta_b} = \sum_{\alpha=1}^N \frac{1}{m_\alpha}
    \frac{\partial \theta_a}{\partial r_{\alpha j}} \frac{\partial
      \theta_b}{\partial r_{\alpha j}} =
    \N^{-1}_{mn}\Lambda^{-1}_{ma}\Lambda^{-1}_{nb}~, \quad
    \N^{-1}_{ij} = \sum_{c,d=1}^3 \R^2_{cd} \Q^{-1}_{ic} \Q^{-1}_{jd}
    = \sum_{\alpha=1}^N \frac{1}{m_\alpha} \F_{i\alpha k} \F_{j\alpha
      k}~.
  \end{gathered}
\end{equation}
The remaining block $k^{-1}_{\theta_a b}$ is obtained from
$k^{-1}_{a\theta_b}$ by symmetry.  The coefficients (\ref{eq:blocks1})
fix the form of $\K_{\theta1}$.  Furthermore, we can eliminate
$\pth{a}$ in favor of the body-frame angular momentum $\mat{L}$ with
the aid of (\ref{eq:angmom}).  Taking account of the ordering of
operators we get,
\begin{equation}
  \label{eq:Ktheta1}
  \begin{aligned}
  \K_{\theta1} &= \sum_{a=1}^{3N-6} \D_{ak} p_a L_k + \frac{1}{2} 
  \N^{-1}_{ij} L_i L_j + \N^{-1}_{ij} \sum_{a,b=1}^3 \Lambda^{-1}_{ia} 
  \frac{\partial \Lambda^{-1}_{jb}}{\partial \theta_a}
  \frac{\partial}{\partial \theta_b}~,\\
  \D_{ak} &= \sum_{\alpha=1}^N \frac{1}{m_\alpha} \F_{k\alpha
  l}\frac{\partial t_a}{\partial R_{\alpha l}}~.
  \end{aligned}
\end{equation}
The coefficients of the first two terms in $\K_{\theta1}$ depend only
on $\{t_a\}$.  There still is dependence on $\{\theta_b\}$ in the last
term, which will cancel against an analogous term in $\K_{\theta2}$.
We turn to the latter next.

The operator $\K_{\theta2}$ is linear in $\pth{a}$, with coefficients
$k_{\theta_a}$ given by (\ref{eq:standard}) with $q_b$ substituted by
$\theta_a$.  The second derivative of $\theta_a$ is obtained by
differentiating both sides of (\ref{eq:thetafund}) with respect to
$r_{\alpha j}$.  Using
\begin{equation*}
  \frac{\partial\Lambda^{-1}_{ma}}{\partial r_{\alpha j}} =
  \sum_{d=1}^3 
  \frac{\partial\theta_d}{\partial r_{\alpha j}}
  \frac{\partial\Lambda^{-1}_{ma}}{\partial \theta_d}
\end{equation*}
with $\partial\theta_d/\partial r_{\alpha j}$ given by
(\ref{eq:thetafund}), and the expression for $\partial U_{ij}/\partial
r_{\alpha j}$ from (\ref{eq:Ufund}), we get,
\begin{equation}
  \label{eq:claux}
  \begin{aligned}
    k_{\theta_a} &= \sum_{\alpha=1}^N \frac{1}{m_\alpha}
    \frac{\partial^2\theta_a}{\partial r_{\alpha j}\partial r_{\alpha
    j}} = 
    \sum_{d=1}^3 \N^{-1}_{lm} \Lambda^{-1}_{ld}
    \frac{\partial\Lambda^{-1}_{ma}}{\partial\theta_d} +
    \Lambda^{-1}_{la} \B_l\\
    \B_l &= -\sum_{\alpha=1}^N \frac{1}{m_\alpha}
    \frac{\partial\F_{l\alpha i}}{\partial r_{\alpha j}} U_{ij} -
    \varepsilon_{inm} \sum_{\alpha=1}^N \frac{1}{m_\alpha} \F_{l\alpha 
      i} \F_{m\alpha n}~.
  \end{aligned}
\end{equation}
Since $\F_{r\alpha i}$ by its definition (\ref{eq:sprep2}) depends on
$\{t_a\}$ only, using the rotation invariance of $t_a$ we can write,
\begin{subequations}
\begin{align}
  \label{eq:clfund}
  \B_l &= -\sum_{\alpha=1}^N \frac{1}{m_\alpha} \left(
    \frac{\partial\F_{l\alpha i}}{\partial R_{\alpha i}} +
    \varepsilon_{inm} \F_{m\alpha n} \F_{l\alpha i}\right) ~,\\
  \intertext{with}~
  \frac{\partial\F_{l\alpha i}}{\partial R_{\alpha i}} &\equiv
  \frac{\partial\F_{l\alpha i}}{\partial r_{\alpha i}}(\{\mat{R}\}) =
    \sum_{a=1}^{3N-6} \frac{\partial\F_{l\alpha i}}{\partial t_a}
    \frac{\partial t_a}{\partial R_{\alpha i}}~. \label{eq:32b}
\end{align}
\end{subequations}
(In appendix \ref{sec:app0} we rewrite (\ref{eq:32b}) in a completely 
different way in terms of \mat{U}.)  Therefore,
\begin{equation}
  \label{eq:Ktheta2}
  \K_{\theta2} = -\frac{1}{2}\sum_{a=1}^3 k_{\theta_a}
  \frac{\partial}{\partial\theta_a} = 
  -\frac{1}{2}\sum_{a=1}^3 \N^{-1}_{lm}
  \Lambda^{-1}_{ld} \frac{\partial
  \Lambda^{-1}_{ma}}{\partial\theta_d}
  \frac{\partial}{\partial\theta_a}+ \frac{i}{2} \B_k L_k ~.
\end{equation}
Thus, $\Kz$, $\K_{\theta1}$ and $\K_{\theta2}$ add up to the total
kinetic energy operator,
\begin{equation}
  \label{eq:Kstand}
  \K = \Kz + \Ko + \Kt~,
  \quad
   \Ko =  \sum_{a=1}^{3N-6} \D_{ak} p_a L_k
   + \frac{i}{2} \B_k L_k ~,
   \quad
    \Kt = \frac{1}{2} \N^{-1}_{ij} L_i L_j   ~,
\end{equation}
where \Kz\ is given in (\ref{eq:kostan}), $\D_{ak}$ and $\B_{k}$ are
given in (\ref{eq:Ktheta1}) and (\ref{eq:clfund}) resp., and
$\N^{-1}_{ij}$ in (\ref{eq:blocks1}).  The notation used in
(\ref{eq:Kstand}) is such that $\K_n$, $n=0,1,2$, depends on the
$n^{\mathrm{th}}$ power of the body-frame angular momentum.  The
coefficient functions in \K\ depend on $\{t_a\}$ only, as expected by
rotation invariance.  Clearly, \K\ is invariant under frame-conditions
reparameterizations, and independent of the parameterization of the
rotation group used to define the orientational variables
$\{\theta_a\}$.  The second term in \Ko\ is purely quantum mechanical,
since $\B_k\propto\hbar$ as can easily be checked by dimensional
analysis.  The origin of this term lies in the operator ordering, in
the same way as other orderings (such as e.g., Weyl ordering) give
rise to quantum potentials \cite{bou04}.  The form of $\B_k$ is
further simplified by the fact that the first term in the parentheses
in (\ref{eq:clfund}) vanishes for the most usual choices of frame such
as the $N$-body Eckart frame, and also in the examples in sections
\ref{sec:examp} and \ref{sec:examp4} below.

\section{Inner product. Matrix and rigid-rotator Hamiltonians}
\label{sec:matrig}

Due to translation invariance the lab-frame wave function can be
factored as $\Psi_{(0)}(\{\mat{r}\}) =
\psi_{(0)}(\{\mat{r}_\alpha-\vrcm\})\exp(i\mat{k}_\mathrm{cm}\cdot\vrcm)$, 
with the subindex (0) indicating lab frame.  Starting with the canonical
inner product in the lab frame and changing variables to $\{t_a\}$,
$\{\theta_b\}$, \vrcm\, we get,
\begin{equation}
  \label{eq:inner}
  \langle\widetilde{\Psi} |\Psi\rangle = \int \prod_{a=1}^{3N-6}dt_a
  \prod_{b=1}^{3}d\theta_b \J
  \widetilde{\psi}_{(0)}^*(\{\mat{U}^\dagger\mat{R}_\alpha\})
  \psi_{(0)}(\{\mat{U}^\dagger\mat{R}_\alpha\}) ~.
\end{equation}
Here we already integrated over \vrcm, obtaining a
momentum-conservation $\delta$ function which we omit. The Jacobian
$\J$ can be expressed in terms of internal coordinates by means of the
relation
\begin{equation}
  \label{eq:detqr}
  \J = \left|\det\left(\frac{\partial q_a}{\partial r_{\alpha
   j}}\right)^{-1}\right| =   
   \left|\det\left(\frac{\partial q_a}{\partial R_{\alpha
   j}}\right)^{-1}\right| =  |\Lambda| \widetilde\J~,
\end{equation}
where $|\Lambda| = \det(\Lambda_{ai})$, with $\Lambda_{ai}$ defined in
(\ref{eq:Ufund}), and $\partial q_a/\partial R_{\alpha j} = U_{kj}
\partial q_a/\partial r_{\alpha k}$ and we used (\ref{eq:thetafund}).
$1/\widetilde\J$ in (\ref{eq:detqr}) is the absolute value of the
determinant of the $3N\times3N$ matrix
\begin{equation}
  \label{eq:detqr1}
  \left(\begin{array}{cccc}
   \partial t_1/\partial R_{1 X} & \partial t_1/\partial R_{1
   Y} & \ldots & \partial t_1/\partial R_{N Z} \\
   \vdots & \vdots &  & \vdots\\
   \partial t_{3N-6}/\partial R_{1X} & \partial t_{3N-6}/\partial
   R_{1Y} &  \ldots& \partial t_{3N-6}/\partial R_{NZ} \\
   -\F_{11X} & -\F_{11Y} & \ldots & -\F_{1NZ}\\
   \vdots & \vdots &  & \vdots\\
   -\F_{31X} & -\F_{31Y} & \ldots & -\F_{3NZ}\\
  \partial r_{\mathrm{cm}X}/\partial r_{1X} & \partial r_{\mathrm{cm}X}/\partial
  r_{1Y} & \ldots & \partial r_{\mathrm{cm}X}/\partial r_{NZ}\\
   \vdots & \vdots &  & \vdots\\
  \partial r_{\mathrm{cm}Z}/\partial r_{1X} & \partial r_{\mathrm{cm}Z}/\partial
  r_{1Y} & \ldots & \partial r_{\mathrm{cm}Z}/\partial r_{NZ}\\
  \end{array}\right)~.
\end{equation}
The minus signs in the three middle rows are of course unimportant in
(\ref{eq:detqr}).  Notice that the quantities $\F_{i\alpha j}$ are
completely determined by the frame conditions, and are usually much
simpler in form than gradients of Euler angles.  The Jacobian $J$ as
given in (\ref{eq:covariant}), which is proportional to $\J$, can be
computed with the derivatives (\ref{eq:thetafund}) in terms of the
matrix $g_{ab}$ from (\ref{eq:kostan}).  The procedure in this case is
closely analogous to the case of linear frame conditions discussed in
\cite{bou04}.  We will not dwell on that calculation, whose result and
its derivation have been considered in \cite{lit97}. With our notation
we have,
\begin{equation}
  \label{eq:jaco}
  J= \prod_{\alpha=1}^N m_\alpha^{3/2} \J=
  M^{3/2} |\Lambda|
  \frac{|\mat{M}|^{1/2}}{|g|^{1/2}}~, 
\end{equation}
where $M=\sum_\alpha m_\alpha$, $\mat{M}$ is the body-frame inertia
tensor and $|\mat{M}|$ its determinant and $|g|=\det g_{ab}$ with
$g_{ab}$ from (\ref{eq:kostan}).  Below we denote $dV_\theta =
\prod_{b=1}^{3}d\theta_b |\Lambda|$ the invariant measure on
$\mathrm{SO}(3)$, with total volume $V_\theta=8\pi^2$.
  
Since $\K$ commutes with $(\mat{l}-\mat{l}_\mathrm{cm})^2 = \mat{L}^2$
and $(l_z-l_{\mathrm{cm}z})$ we can choose $\psi_{(0)}=\psi_{(0)\ell
  m}(\{\mat{r}_\alpha - \vrcm\})$ to be an eigenfunction of those
operators.  The body-frame wave functions are then
\begin{subequations}
  \label{eq:bodywave}
\begin{align}
  \psi_{\ell m}(\{\mat{R}\},\{\theta_a\}) &= \psi_{(0)\ell
    m}(\{\mat{U}^\dagger(\{\theta_a\})\mat{R}\}) = \sqrt{2\ell+1} 
  \sum_{s=-\ell}^\ell \psi_{(0)\ell s}(\{\mat{R}\}) D^{\ell
    *}_{ms}(\{\theta_a\}) \\
\intertext{with}
  D^{\ell *}_{m'm}(\{\theta_a\}) &= \int d^2\ver{e}\,
    Y^*_{lm'}(\ver{e}) 
    Y_{lm}(\mat{U}(\{\theta_a\})\ver{e}) 
\end{align}
\end{subequations}
where $\ver{e}$ is a unit vector varying over the unit sphere and
$D^{\ell}_{m'm}(\{\theta_a\})$ the irreducible matrix representing the
rotation $\mat{U}(\{\theta_a\})$.  In terms of the wave functions
(\ref{eq:bodywave}) we have,
\begin{equation}
  \label{eq:inner2}
    \langle\widetilde{\Psi}_{\ell m} |\Psi_{\ell m'}\rangle =
    \delta_{mm'}V_\theta\sum_{n=-\ell}^\ell\int
    \prod_{a=1}^{3N-6}dt_a\, 
    \widetilde{\J} \widetilde{\psi}_{(0)_{\ell
    n}}^*(\{\mat{R}\}) 
  \psi_{(0)_{\ell n}}(\{\mat{R}\}) ~. 
\end{equation}
The action of the lab frame angular momentum on the body-frame wave
functions (\ref{eq:bodywave}) is given by,
\begin{equation}
  \label{eq:laban}
  l_i \psi_{\ell m}(\{\mat{R}\},\{\theta_a\}) = -\sqrt{2\ell+1}
  \sum_{s,q=-\ell}^\ell 
  \psi_{(0)\ell s}(\{\mat{R}\}) \left(\Ll^{(\ell)}_i\right)_{qm}
  D^{\ell*}_{qs}(\{\theta_a\})~,
\end{equation}
where $\Ll^{(\ell)}_i$ is the standard angular momentum Hermitian
matrix in the representation of irreducible tensors of order $\ell$,
\cite{brn}
\begin{equation}
  \label{eq:irredmat}
  \left(\Ll^{(\ell)}_i\right)_{km} = \int d^2 \ver{e}\,
  Y^*_{lk}(\ver{e}) \varepsilon_{ipq} e_p
  \frac{1}{i}\frac{\partial}{\partial e_q} Y_{lm}(\ver{e})
  = \sum_{a=1}^3 \left(\lambda^{-1}_{ia}(\{\alpha_a\}) \frac{1}{i}
  \frac{\partial}{\partial\alpha_a}\right)_{\alpha_a=0}
  D^\ell_{km}(\{\alpha_a\}) ~.
\end{equation}
The matrix $\lambda^{-1}_{ia}(\{\alpha_a\})$ in (\ref{eq:irredmat}) is
as defined in (\ref{eq:lambdas}).
Analogously, the body-frame angular momentum operator acts as,
\begin{equation}
  \label{eq:bodan}
  L_i \psi_{\ell m}(\{\mat{R}\},\{\theta_a\}) = -\sqrt{2\ell+1} 
  \sum_{s,q=-\ell}^\ell 
  \psi_{(0)\ell s}(\{\mat{R}\}) \left(\Ll^{(\ell)}_i\right)_{sq}
  D^{\ell*}_{mq}(\{\theta_a\})~.
\end{equation}
Thus, we can represent \K\ by means of its matrix elements between
angular-momentum eigenfunctions in terms of the matrices
$\Ll^{(\ell)}_i$.  We need consider only matrix elements between wave
functions with different ``radial'' quantum numbers, but the same
angular dependence, so that,
\begin{equation}
  \label{eq:matham}
  \begin{gathered}
    \frac{1}{V_\theta}
    \int dV_\theta\,\widetilde{\psi}^*_{lm}(\{\mat{R}\},\{\theta_a\})
    \K \psi_{lm}(\{\mat{R}\},\{\theta_b\}) = \sum_{p,q=-\ell}^\ell
    \widetilde{\psi}^*_{(0)lp}(\{\mat{R}\}) \widehat{\K}_{pq}
    \psi_{(0)lq}(\{\mat{R}\}) \\
    \widehat{\K}_{pq} \equiv \Kz\delta_{pq} - \sum_{a=1}^{3N-6} \D_{ak}
    p_a \left(\Ll^{(\ell)}_k\right)_{qp} - \frac{i}{2} \B_k
    \left(\Ll^{(\ell)}_k\right)_{qp} + \frac{1}{2} \N^{-1}_{ij}
    \left(\Ll^{(\ell)}_i\right)_{qr}
    \left(\Ll^{(\ell)}_j\right)_{rp}~.  
  \end{gathered}
\end{equation}
In this equation the operator \Kz\ and the coefficient functions
$\D_{ak}$, $\B_k$ and $\N^{-1}_{ij}$ are as in (\ref{eq:Kstand}).

Instead of the body frame wave functions (\ref{eq:bodywave}) we can
introduce the alternative basis,
\begin{equation}
  \label{eq:rotorwave}
  \phi_\ell(\{\mat{R}\},\ver{e}) = \sum_{m=-\ell}^\ell
  \psi_{(0)lm}(\{\mat{R}\}) Y_{lm}^*(\ver{e})~.
\end{equation}
These wave functions, not eigenfunctions of $l_z$, depend on a unit
vector $\ver{e}$ representing a fictitious rigid rotator of total
angular momentum $\ell$.  In terms of $\phi_\ell$ we have,
\begin{subequations}
  \begin{gather}
    \frac{1}{V_\theta}
  \int dV_\theta\,\widetilde{\psi}^*_{lr}(\{\mat{R}\},\{\theta_a\})
    L_p \psi_{ls}(\{\mat{R}\},\{\theta_b\}) = \delta_{rs} \int d^2
    \ver{e}\,\widetilde{\phi}_\ell^*(\{\mat{R}\},\ver{e}) S_p
    \phi_\ell^*(\{\mat{R}\},\ver{e}) ~,  \label{eq:rotangmom}\\
    S_p = \frac{1}{i} \varepsilon_{pqr} e_q \frac{\partial}{\partial
    e_r}~. \label{eq:spino}
  \end{gather}
\end{subequations}
The factor $\delta_{rs}$ on the r.h.s.\ of (\ref{eq:rotangmom}) is due
to the fact that $L_p$ commutes with $l_z$ (see (\ref{eq:commute})).
Expression (\ref{eq:spino}) for $S_p$ is not affected by the
constraint $\ver{e}\cdot\ver{e} = 1$, that is, the derivatives can be
computed without taking that constraint into account, as can be easily
checked.  Alternatively, the operator $S_p$ can be expressed in terms
of the spherical angles $\theta$, $\phi$ of \ver{e}\ and derivatives
with respect to them.  Therefore, from (\ref{eq:Kstand}) and
(\ref{eq:rotangmom}) we have
\begin{equation}
  \label{eq:rotham}
  \frac{1}{V_\theta}
 \int dV_\theta \widetilde{\psi}^*_{lm}(\{\mat{R}\},\{\theta_a\})
   \K \psi_{lm}(\{\mat{R}\},\{\theta_b\}) = 
 \int d^2\ver{e} \widetilde{\phi}^*_{l}(\{\mat{R}\},\ver{e})
   \widehat{\K} \phi_{l}(\{\mat{R}\},\ver{e})
\end{equation}
with $\widehat{\K}$ having the same form as \K\ in (\ref{eq:Kstand}),
but with the angular momentum \mat{L}\ replaced by \mat{S}\ as given
by (\ref{eq:spino}).  Actually, the product $L_iL_j$ in
(\ref{eq:Kstand}) is mapped into $S_jS_i$, but that product is
contracted with  $\N^{-1}_{ij}$ which is symmetric.  This
rigid-rotator formalism, based on the wave functions
(\ref{eq:rotorwave}) and the Hamiltonian (\ref{eq:rotham}), is a useful
alternative to the matrix formalism based on (\ref{eq:bodywave}),
(\ref{eq:irredmat}) and (\ref{eq:matham}).  It appears naturally in
the gauge-invariant approach of \cite{bou04}.

\section{Other forms for the Hamiltonian}
\label{sec:other}

Other forms for the many-body Hamiltonian in a body-fixed frame, based
on expression (\ref{eq:covariant}) for the Laplacian, have been given
in the literature (see, e.g., \cites{wat68,lit97} and references
therein).  In this section we discuss the derivation of the kinetic
energy operator in the form (\ref{eq:covariant}) from the point of
view of frame conditions and establish relations among these results
and those of section \ref{sec:stand}.  Our notation follows 
that of \cite{lit97}.

In order to express the kinetic energy in the form
(\ref{eq:covariant}) it is convenient to write the matrix
$k^{-1}_{ab}$ in a form different from that used in sec.\ 
\ref{sec:stand}.  Defining,
\begin{equation}
  \label{eq:hab}
  h_{ab} = \sum_{\alpha=1}^N m_\alpha \frac{\partial R_{\alpha
  i}}{\partial t_a} \frac{\partial R_{\alpha i}}{\partial t_b}
\end{equation}
and using the chain rule we get,
\begin{equation}
  \label{eq:lemma1-p15}
  \frac{\partial t_a}{\partial r_{\alpha i}} = \sum_{b=1}^{3N-6}
  \sum_{\beta=1}^N m_\beta h^{-1}_{ab} \frac{\partial R_{\alpha
  j}}{\partial t_b} \frac{\partial R_{\alpha j}}{\partial r_{\alpha
  i}} ~.
\end{equation}
Substituting (\ref{eq:lemma1-p15}) in the definition (\ref{eq:kostan})
of $g_{ab}$, and using the derivatives (\ref{eq:dR/dr2}), we get,
\begin{equation}
  \label{eq:lemma2-p15}
  g_{ab} = h^{-1}_{ab} + \sum_{c,d=1}^{3N-6} h^{-1}_{ac} h^{-1}_{bd}
  a_{ci}a_{dj} \N^{-1}_{ij}
\end{equation}
with $\N^{-1}_{ij}$ defined in (\ref{eq:blocks1}) and
\begin{equation}
  \label{eq:a}
  \mat{a}_d = \sum_{\beta=1}^N m_\beta
  \mat{R}_\beta\wedge\frac{\partial\mat{R}_\beta}{\partial t_a}~.
\end{equation}
Several relations between $g_{ab}$ and $h_{ab}$, and between \mat{M}\
and \mat{\N}, analogous to (\ref{eq:lemma2-p15}) are summarized in
appendix \ref{sec:appa}.  Eq.\ (\ref{eq:lemma2-p15}) fixes the form of
$k^{-1}_{ab}$ in (\ref{eq:covariant}) for $1\leq a,b\leq 3N-6$.

Similarly, we can obtain a compact expression for the off-diagonal
blocks of $k^{-1}_{ab}$.  Using (\ref{eq:lemma1-p15}) and
(\ref{eq:dR/dr2}) together with the frame conditions $\G_a$ and their
translation invariance, we obtain,
\begin{equation}
  \label{eq:blocks2}
  k^{-1}_{a\theta_b} = \sum_{c=1}^{3N-6} h^{-1}_{ac} a_{cj}
  \N^{-1}_{jk} \Lambda^{-1}_{kb}~, 
\end{equation}
to be compared with the corresponding expression in
(\ref{eq:blocks1}).  The block $k^{-1}_{\theta_a\theta_b}$ is as given
in (\ref{eq:blocks1}).  As in sect.\ \ref{sec:stand} we omit here for
brevity the terms involving the center-of-mass degrees of freedom,
which are dynamically trivial. 
With these expressions for $k^{-1}_{ab}$ we can compute its
determinant $1/J^2$, by factoring the matrix appropriately.  We again
omit the details \cites{lit97,bou04} and state the result 
\begin{equation}
  \label{eq:jaco2}
  J=M^{3/2} |\Lambda| |\N|^{1/2} |h|^{1/2}~,
\end{equation}
with $|h|=\det(h_{ab})$ and $|\N|=\det(\N)$.  Comparing
(\ref{eq:jaco2}) with (\ref{eq:jaco}) 
yields $|\mat{M}|/|g| = |h| |\N|$ \cite{lit97}.

The matrix $k^{-1}_{ab}$ and $J$ are all we need in order to obtain
the kinetic-energy operator \K\ from (\ref{eq:covariant}).  We can,
however, eliminate all dependence on orientational degrees of freedom
by means of the well-known relations (see \cite{bou04} and refs.\
therein) 
\begin{equation}
  \label{eq:angmomaux}
  L_i = i\sum_{a=1}^3 \Lambda^{-1}_{ia}
  \frac{\partial}{\partial\theta_a} = \frac{i}{|\Lambda|}\sum_{b=1}^3
  \frac{\partial}{\partial\theta_b} \Lambda^{-1}_{ib} |\Lambda|~.
\end{equation}
With this, we finally get,
\begin{equation}
  \label{eq:covariant2}
  \begin{aligned}
    \K &= \frac{1}{2|\N|^{1/2} |h|^{1/2}} \sum_{a,b=1}^{3N-6} p_{a}
    h^{-1}_{ab} |h|^{1/2} |\N|^{1/2} p_{b}\\
    &\mbox{ } + \frac{1}{2|\N|^{1/2} |h|^{1/2}}
    \left(L_i - \sum_{b,b'=1}^{3N-6} p_{b} h^{-1}_{bb'}
    a_{b'i}\right) |\N|^{1/2} |h|^{1/2} \N^{-1}_{ij}
    \left(L_j - \sum_{d,d'=1}^{3N-6} h^{-1}_{dd'}
    a_{d'j} p_{d} \right) ~.
  \end{aligned}
\end{equation}
Notice the ordering of operators in (\ref{eq:covariant2}).  \K\ can be
Weyl ordered most easily after performing a transformation of the form
$J\K\ 1/J$, leading to a quantum potential term.  Weyl ordering is
considered in detail in the case of linear frame conditions in
\cite{bou04}, and the same procedure can be applied to the case of
general frame conditions discussed in this paper. The expression for
the quantum potential, however, seems to us to be too complicated to
be useful in practice so we omit the results.  Other orderings are of
course possible, although with similar caveats about the associated
quantum potentials.  That is an advantage, from our point of view, of
the standard ordering given in section \ref{sec:stand}.

We can rewrite \K\ as given by (\ref{eq:covariant2}) in terms of
$g_{ab}$ and $\mat{M}^{-1}$, instead of $h^{-1}_{ab}$ and
$\mat{\N}^{-1}$, by using relations (\ref{eq:lemma2-p15}),
(\ref{eq:jaco}) and (\ref{eq:jaco2}), and (\ref{eq:corollary2-p16.1}),
to find 
\begin{equation}
  \label{eq:covariant3}
  \K = \frac{1}{2} M_{ij}^{-1} L_iL_j + \frac{1}{2}
  \frac{|g|^{1/2}}{|M|^{1/2}} \sum_{a,b=1}^{3N-6} \left( p_a -
  A_{ai}L_i\right) \frac{|M|^{1/2}}{|g|^{1/2}} g_{ab} \left(p_b -
  A_{bj}L_j \right)~.
\end{equation}
Here we defined \cite{lit97}
\begin{equation}
  \label{eq:Aa}
  \mat{A}_a = \mat{M}^{-1} \mat{a}_a~,
  \qquad 1\leq a\leq 3N-6~.
\end{equation}
In the form (\ref{eq:covariant3}) all dependence of \K\ on frame
conditions is implicit in the relations
$\mat{R}_\alpha=\mat{R}_\alpha(\{t_a\})$, which enters \K\ through
$g_{ab}$ and $\mat{A}_a$ and also through $\mat{M}^{-1}$ when
expressed in terms of internal coordinates.  It is interesting to
point out that for the most commonly used frames the 3$\times 3$
matrix $\Q_{ai}$ defined in (\ref{eq:frnot}) is much simpler to invert
than \mat{M}, and that is the only matrix inversion needed to obtain
\K\ as given in (\ref{eq:Kstand}).

Comparing the expressions (\ref{eq:covariant2}) and
(\ref{eq:covariant3}) with (\ref{eq:Kstand}) we can obtain relations
among their coefficients.  The equivalence of the purely vibrational
terms in (\ref{eq:Kstand}) and (\ref{eq:covariant2}) is immediate once
we take into account (\ref{eq:lemma6-p18}) and the equivalence between
the two standard forms for the Laplacian (\ref{eq:standard}) and
(\ref{eq:covariant}).  Similarly, the terms quadratic in \mat{L}\ in
(\ref{eq:Kstand}) and (\ref{eq:covariant2}) are obviously equal and
equivalent to that in (\ref{eq:covariant3}) by
(\ref{eq:corollary2-p16.1}).  Notice that $\N^{-1}_{ij}$, which is
usually defined \cite{lit97} as in (\ref{eq:lemma3-p16}) or
(\ref{eq:corollary2-p16.1}), can be compactly expressed in terms of
frame conditions by our definition (\ref{eq:blocks1}).

Equating the terms linear in \mat{L}\ in (\ref{eq:Kstand}),
(\ref{eq:covariant2}) and (\ref{eq:covariant3}) we get the relations,
\begin{subequations}
\label{eq:relin}
\begin{align}
  \D_{dq} &= -M^{-1}_{qr}\sum_{d'=1}^{3N-6} g_{dd'} a_{d'r} =
  -\N^{-1}_{qr} \sum_{d'=1}^{3N-6} h^{-1}_{dd'} a_{d'r}   \label{eq:relina}\\
  \frac{i}{2} \B_q &= \frac{1}{2} \frac{|g|^{1/2}}{|M|^{1/2}}
  \sum_{d=1}^{3N-6} \left(p_{d}
    \frac{|M|^{1/2}}{|g|^{1/2}} \D_{dq}\right)
  = \frac{1}{2 |\N|^{1/2}|h|^{1/2}} \sum_{d=1}^{3N-6} \left(
  p_d |\N|^{1/2}|h|^{1/2} \D_{dq}\right)
  \label{eq:relinb}
\end{align}
\end{subequations}
with $\D_{dq}$ and $\B_q$ defined in (\ref{eq:Ktheta1}) and
(\ref{eq:claux}), respectively.  These relations can be proved
directly, providing a consistency check on our results. 
An important consequence of (\ref{eq:relina}) is that it allows us to
write $\mat{A}_a$, at least locally, in terms of
frame conditions,
\begin{equation}
  \label{eq:A}
  A_{bi} = -\sum_{c=1}^{3N-6} g^{-1}_{bc} \D_{ci}~,
  \qquad\text{or}\qquad
 a_{bi} = -\N_{ik}\sum_{c=1}^{3N-6} h_{bc} \D_{ci}~,  
\qquad a=1,\ldots,3N-6~,
\end{equation}
with $\D_{ci}$ given by (\ref{eq:Ktheta1}).  Notice that these
relations cannot be obtained from (\ref{eq:lemma2-p15}) or the
equalities in appendix \ref{sec:appa}, which always involve
$\mat{a}_a$ or $\mat{A}_a$ quadratically.  Through (\ref{eq:A}) we can
write any expression involving the gauge fields $\mat{A}_a$
\cite{lit97} in terms of internal coordinates and frame
conditions.

\section{The case \mat{N}=3}
\label{sec:examp}

We consider here the case $N=3$ both as an example and a verification
of the foregoing, obtaining the Hamiltonian in two different body
frames, one defined by linear conditions and the other by quadratic
ones.  We choose internal coordinates $t_1\equiv\rho_1$,
$t_2\equiv\rho_2$ and $t_3\equiv\theta$ which are standard in
molecular physics,
\begin{equation}
  \label{eq:3coord}
  \rho_1 = |\mat{r}_1-\mat{r}_3|~,\quad
  \rho_2 = |\mat{r}_2-\mat{r}_3|~,\quad
  \cos\theta = \frac{1}{\rho_1\rho_2} (\mat{r}_1-\mat{r}_3)\cdot
  (\mat{r}_2-\mat{r}_3)~. 
\end{equation}
with conjugate momenta denoted by $p_a$, $a=1,2,3$. We define also the
reduced masses $1/m_{13}=1/m_1+1/m_3$ and analogously $m_{23}$.

\subsection{Linear frame conditions}
\label{sec:linfram}

A linear body frame with origin at the center of mass can be defined
by choosing the $Y$ axis orthogonal to the plane of the system,
$\ver{Y}\propto (\mat{r}_2-\mat{r}_3) \wedge (\mat{r}_1-\mat{r}_3)$,
the $Z$ axis along $\mat{r}_1-\mat{r}_3$, and
$\ver{X}=\ver{Y}\wedge\ver{Z}$.  The frame conditions are then,
\begin{equation}
  \label{eq:linfram}
  \mat{\C}\equiv \sum_{\alpha=1}^N \frac{m_\alpha}{M} \mat{r}_\alpha
  =0~,
  \quad
  \G_1\equiv r_{1x}-r_{3x}=0~,
  \quad
  \G_2\equiv r_{1y}-\C_{y}=0~,
  \quad
  \G_3\equiv r_{2y}-\C_{y}=0~.
\end{equation}
Notice that $\G_a$ are written so they are explicitly translation
invariant.  
From (\ref{eq:kostan}) and (\ref{eq:3coord}) we get,
\begin{equation}
  \label{eq:gabgb}
  \begin{aligned}
    g_{11} &= \frac{1}{m_{13}}~, & g_{12} &= \frac{\cos\theta}{m_3}~,
    & g_{13} &= -\frac{\sin\theta}{m_3\rho_2}\\
    g_{22} &= \frac{1}{m_{23}}~, & g_{23} &=
    -\frac{\sin\theta}{m_3\rho_1}~, & g_{33} &=
    \frac{1}{m_{13}\rho^2_1}+\frac{1}{m_{23}\rho^2_2} -
    \frac{2\cos\theta}{m_{3}\rho_1\rho_2}\\
    g_1 &= \frac{2}{m_{13}\rho_1}~, & g_2 &= \frac{2}{m_{23}\rho_2}~,
    & g_3 &= \cot\theta\left(\frac{1}{m_{13}\rho^2_1} +
      \frac{1}{m_{23}\rho^2_2}\right) - \frac{2}{m_{3}\rho_1\rho_2
      \sin\theta}~,
  \end{aligned}
\end{equation}
and the Jacobian entering the inner product (\ref{eq:inner2}) is found
to be $\widetilde{\J}=\rho^2_1\rho^2_2\sin\theta$.  The coefficients
(\ref{eq:gabgb}) fix the form of \Kz\ as given in (\ref{eq:standard}).
With the frame conditions (\ref{eq:linfram}) and the internal
coordinates (\ref{eq:3coord}), from (\ref{eq:sprep2}) we obtain,
\begin{equation}
  \label{eq:flin}
  \begin{aligned}
    \mat{\F}_{11} &= -\frac{1}{\rho_1}\ver{Y}~,
    &\mat{\F}_{12} &= 0~,
    &\mat{\F}_{13} &= \frac{1}{\rho_1}\ver{Y}~,\\
    \mat{\F}_{21} &= \frac{1}{\rho_1}\ver{X}~,   
    &\mat{\F}_{22} &= 0~,
    &\mat{\F}_{23} &= -\frac{1}{\rho_1}\ver{X}~,\\
    \mat{\F}_{31} &= -\frac{\cot\theta}{\rho_1} \ver{Y}~,
    &\mat{\F}_{32} &= \frac{1}{\rho_2\sin\theta} \ver{Y}~,    
    &\mat{\F}_{33} &= -\frac{\rho_1 -
    \rho_2\cos\theta}{\rho_1\rho_2\sin\theta} \ver{Y}~, 
  \end{aligned}
\end{equation}
with the notation $\mat{\F}_{a\alpha}\equiv(\F_{a\alpha 1},\F_{a\alpha
  2},\F_{a\alpha 3})$.  We omit the details of the calculation of
$\N^{-1}_{ij}$, $\D_{ak}$ and $\B_l$ (see (\ref{eq:blocks1}),
(\ref{eq:Ktheta1}) and (\ref{eq:claux}), resp.).  Rather, we give the
result for \K, from which those coefficients can be read off.  The
kinetic-energy operator for this system is given by (\ref{eq:Kstand})
as $\K=\Kz+\Ko+\Kt$, with \Kz\ resulting from (\ref{eq:gabgb}) and
with
  \begin{equation}
    \label{eq:Kstand3lin}
    \begin{aligned}
      \Ko &= -\frac{\sin\theta}{m_3\rho_1} p_2L_Y +
      \left(\frac{1}{m_{13}\rho_1^2} -
        \frac{\cos\theta}{m_3\rho_1\rho_2}\right) p_3L_Y +
      \frac{i}{2\sin\theta} \left(\frac{1}{m_3\rho_1\rho_2} -
        \frac{\cos\theta}{m_{13}\rho_1^2}\right)L_Y\\
      \Kt &= \frac{1}{2m_{13}\rho_1^2}L^2_X +
      \frac{1}{2}\left(\frac{1}{m_3\rho_1\rho_2\sin\theta} -
        \frac{1}{m_{13}\rho_1^2\tan\theta} \right)\{L_X,L_Z\} +
      \frac{1}{2m_{13}\rho_1^2}L_Y^2\\
      &\mbox{ }+ \frac{1}{2} \left(\left(\frac{1}{m_{13}\rho^2_1}
          +\frac{1}{m_{23}\rho^2_2} \right)\frac{1}{\sin^2\theta} -
        \frac{1}{m_{13}\rho^2_1} -
        \frac{2\cos\theta}{m_3\rho_1\rho_2\sin^2\theta}\right) L^2_Z
      ~.
    \end{aligned}
  \end{equation}
These results agree exactly with those of \cite{han87} once we take
into account that the kinetic operator defined there is
$\rho_1\rho_2\K\, 1/(\rho_1\rho_2)$ in our notation.  

In this example, since the frame conditions are linear, we can choose
a set of linear body-frame coordinates satisfying (\ref{eq:add1}).  We
set $Q_1=R_{1Z}-R_{3Z}$, $Q_2=R_{2Z}-R_{3Z}$, $Q_3=R_{2X}-R_{3X}$.
These coordinates $Q_a$ can be extended to all of
configuration space by linearity, yielding a set of non-rotation-%
invariant coordinates $Q_1(\{\mat{r}\})=r_{1z}-r_{3z}$, etc.  In order
to extend them to rotation-invariant internal coordinates we express
them in terms of scalar products of body-frame position-vectors.
Such procedure leads to a set of coordinates equivalent to
(\ref{eq:3coord}), which with the same notation are written as
$\rho_1$, $\rho_2\cos\theta$, $\rho_2\sin\theta$.

\subsection{Quadratic frame conditions}
\label{sec:quadfram}

Another frame for the three-body system used in the molecular-physics
literature is defined as a modification of the previous one, choosing
the $Z$ axis to bisect the angle $\theta$ between
$\mat{r}_1-\mat{r}_3$ and $\mat{r}_2-\mat{r}_3$.  The frame conditions
are as in (\ref{eq:linfram}), except that $\G_1$ now takes the form
\begin{equation}
  \label{eq:quadfram}
  \G_1(\{\mat{r}\}) \equiv
  (r_{1x}-r_{3x})(r_{2z}-r_{3z})+(r_{1z}-r_{3z})(r_{2x}-r_{3x}) =0~.
\end{equation}
This frame differs from the body-frame of section \ref{sec:linfram} by
a time-dependent rotation in an angle $\theta/2$ around the \ver{Y}\ 
axis.
Since internal coordinates are rotation invariant, $g_{ab}$ and $g_b$,
and therefore also \Kz, are as in (\ref{eq:gabgb}).  The Jacobian
$\widetilde{\J}$ also remains the same as above.

The modified frame conditions (\ref{eq:quadfram}) lead to,
\begin{equation}
  \label{eq:fquad}
  \begin{aligned}
  \F_{11} &= -\frac{1}{2\rho_1\cos\frac{\theta}{2}}\ver{Y}~,
  &\F_{21} &= \frac{1}{2\rho_1}\cos\frac{\theta}{2}\ver{X} -
 \frac{1}{2\rho_1}\sin\frac{\theta}{2}\ver{Z}~,
  &\F_{31} &= \frac{1}{2\rho_1\sin\frac{\theta}{2}}\ver{Y}\\
  \F_{12} &= -\frac{1}{2\rho_2\cos\frac{\theta}{2}}\ver{Y}~,
 &\F_{22} &= \frac{1}{2\rho_2}\cos\frac{\theta}{2}\ver{X} +
 \frac{1}{2\rho_2}\sin\frac{\theta}{2}\ver{Z}~,  
 &\F_{32} &= -\frac{1}{2\rho_2\sin\frac{\theta}{2}}\ver{Y}\\
 \F_{13} &= -\frac{\rho_1+\rho_2}{2\rho_1\rho_2\cos\frac{\theta}{2}}
 \ver{Y}~,
 &\F_{23} &= -\frac{1}{2\rho_+} \cos\frac{\theta}{2} \ver{X}
  -\frac{1}{2\rho_-} \sin\frac{\theta}{2} \ver{Z}~,
 &\F_{33} &= \frac{\rho_1-\rho_2}{2\rho_1\rho_2\sin\frac{\theta}{2}}
 \ver{Y}~,
  \end{aligned}
\end{equation}
where the notation is as in (\ref{eq:flin}) and $1/\rho_\pm=\pm
1/\rho_1+1/\rho_2$.  The kinetic energy is then $\K=\Kz+\Ko+\Kt$, with
\Kz\ given by (\ref{eq:gabgb}) and with
\begin{align}
  \label{eq:Kstand3quad}
    \Ko &= \frac{\sin\theta}{2m_3}\left( \frac{1}{\rho_2} p_1 -
      \frac{1}{\rho_1} p_2 \right) L_Y + \frac{1}{2}
      \left(\frac{1}{m_{13}\rho^2_1} - 
        \frac{1}{m_{23}\rho^2_2}\right)
      \left(\frac{\cot\theta}{2i} + p_3\right) L_Y\nonumber\\
      \Kt &= \frac{1}{8\cos^2\frac{\theta}{2}}\left(
      \frac{1}{m_{13}\rho^2_1} + \frac{1}{m_{23}\rho^2_2} +
      \frac{2}{m_{3}\rho_1\rho_2}\right) L^2_X
    +\frac{1}{8}\left(
      \frac{1}{m_{13}\rho^2_1} + \frac{1}{m_{23}\rho^2_2} +
      \frac{2\cos\theta}{m_{3}\rho_1\rho_2}\right) L^2_Y\\
    &\mbox{ } + \frac{1}{8\sin^2\frac{\theta}{2}}\left(
      \frac{1}{m_{13}\rho^2_1} + \frac{1}{m_{23}\rho^2_2} -
      \frac{2}{m_{3}\rho_1\rho_2}\right) L^2_Z +
    \frac{1}{4\sin\theta}\left(
      \frac{1}{m_{23}\rho^2_2} - \frac{1}{m_{13}\rho^2_1}\right)
      \{L_X,L_Z\} ~.\nonumber
\end{align}
This expression for \K\ agrees with the result given in \cite{car83},
as corrected in \cite{han87}, taking into account that their operator
corresponds to $\rho_1\rho_2\K\, 1/(\rho_1\rho_2)$ in our notation.

\section{The case \mat{N}=4}
\label{sec:examp4}

As a further example we consider in this section a four-particle
system.  Our choices of frame and internal coordinates below are
appropriate for a system with the topology of the formaldehyde
molecule, though the results are also applicable to other systems for
which those choices are not singular at the equilibrium configuration.
The vibrational Hamiltonian for the formaldehyde molecule has been
given, in the Born-Oppenheimer approximation, e.g., in \cite{han87}
(see section 4.3 and appendix A.)  Other explicit results for
four-body systems are given in \cite{mer03} and references therein.
Here we use a set of internal coordinates which, combined with the
general results given above, greatly simplify calculations and lead to
moderately simple results for the total Hamiltonian, including
rotation and vibration-rotation terms.  One drawback of our coordinate
choice, however, is that it also results in a complicated expression
for the inner product.  This section is not meant as an exhaustive
kinematic analysis of the four-body problem, but rather as an example
of the results given above.

In this section we label the particles with capital letters, $\alpha =
A,\ldots,D$.  In the case of the formaldehyde molecule $D$ would refer
to the carbon atom, $A$ to the oxygen, and $B$ and $C$ to the hydrogen
atoms. We choose a frame with origin at the center of mass whose $Z$
axis lies along $\mat{R}_{AD}\equiv \mat{R}_{A}-\mat{R}_{D}$, and the
$Y$ axis is defined by the condition that $\mat{R}_{CD}$ lies on the
$YZ$ coordinate plane.  This choice of frame is singular when
$\mat{R}_{CD}$ is parallel to $\mat{R}_{AD}$.  The rotational frame
conditions are,
\begin{equation}
  \label{eq:4frame}
  \G_1\equiv r_{ADx}=0~,
  \quad
  \G_2\equiv r_{ADy}=0~,
  \quad
  \G_3\equiv r_{CDx}=0~.  
\end{equation}
The frame is completely determined by (\ref{eq:4frame}) together with
the auxiliary conditions $R_{AD Z}>0$ and $R_{CD Y}>0$ defining the
direction of the axes.  From (\ref{eq:4frame}) and (\ref{eq:frnot}) we
get, 
\begin{equation}
  \label{eq:4frm1}
  \R^2_{ab} =\left(
  \begin{array}{ccc}
    \mu_{AD}^{-1} & 0 & \mu_{D}^{-1}\\
    0    & \mu_{AD}^{-1} & 0 \\
    \mu_{D}^{-1}  & 0 & \mu_{CD}^{-1}
  \end{array}\right)~,
\qquad
  \Q_{ai} = 
  \left(\begin{array}{ccc}
    0 & R_{ADZ} & 0\\
    -R_{ADZ} & 0 & 0 \\
    0  & R_{CDZ} & -R_{CDY}
  \end{array}\right)~,
\end{equation}
with $1/\mu_{AD} = 1/\mu_{A}+1/\mu_{D}$ and similarly for the other
reduced masses.  The frame conditions are therefore singular when
$\det(\Q)=-R_{ADZ}^2 R_{CDY} =0$, i.e., when $\mat{R}_{AD}$ and
$\mat{R}_{CD}$ are parallel, or either one vanishes.  With the matrix
$\Q$ in (\ref{eq:4frm1}) from (\ref{eq:sprep2}) we obtain
\begin{equation}
  \label{eq:4frm2}
  \begin{gathered}
  \F_{1A2}=-\frac{1}{R_{ADZ}}~,
  \quad
  \F_{1D2}=\frac{1}{R_{ADZ}}~,
  \quad
  \F_{2A1}=\frac{1}{R_{ADZ}}~,
  \quad
  \F_{2D1}=-\frac{1}{R_{ADZ}}~,\\
  \F_{3A1}=\frac{R_{CDZ}}{R_{ADZ}R_{CDY}}~,
  \quad
  \F_{3C1}=-\frac{1}{R_{CDY}}~,
  \quad
  \F_{3D1}=\frac{R_{ADZ}-R_{CDZ}}{R_{ADZ}R_{CDY}}~,
  \end{gathered}
\end{equation}
all other $\F_{i\alpha j}$ vanishing.  In turn this leads to
\begin{equation}
  \label{eq:4frm3}
  \begin{gathered}
    \N^{-1}_{11} = \frac{1}{\mu_{AD}R_{ADZ}^2}~,
    \quad
    \N^{-1}_{23} = \frac{R_{CDZ}}{\mu_{AD}R_{ADZ}^2R_{CDY}} -
    \frac{1}{\mu_DR_{ADZ}R_{CDY} } = \N^{-1}_{32}~,    \\
    \N^{-1}_{22} = \frac{1}{\mu_{AD}R_{ADZ}^2}~,
    \quad
    \N^{-1}_{33} = \frac{R_{CDZ}^2}{\mu_{AD}R_{ADZ}^2R_{CDY}^2}
    + \frac{1}{\mu_{CD}R_{CDY}^2} - 2 \frac{R_{CDZ}}{\mu_DR_{ADZ}R_{CDY}^2 }~, 
  \end{gathered}
\end{equation}
and the remaining components vanishing.  With $\N^{-1}$ from
(\ref{eq:4frm3}), the rotational kinetic energy $\Kt$
(\ref{eq:Kstand}) is completely determined.

Our choice of internal coordinates is motivated by calculational
simplicity.  We introduce translation- and rotation-invariant internal
coordinates depending polynomially on the position vectors,
\begin{equation}
  \label{eq:4frm4}
  \begin{gathered}
  t_1=\mat{r}_{AD}^2~,\quad
  t_2=\mat{r}_{CD}^2~,\quad
  t_3=\mat{r}_{CD}\cdot\mat{r}_{AD}~,\\
  t_4=\mat{r}_{BD}\cdot\mat{r}_{AD}~,\quad
  t_5=\mat{r}_{BD}\cdot\mat{r}_{CD}~,\quad
  t_6=\mat{r}_{BD}\cdot\mat{r}_{CD}\wedge\mat{r}_{AD}~.
  \end{gathered}
\end{equation}
The ranges of variation for the first three coordinates are
$t_{1,2}>0$, $|t_3|<t_1^{1/2} t_2^{1/2}$, whereas the last three can
take any real value.  In the frame defined by conditions
(\ref{eq:4frame}) and the associated suplementary conditions, relative
particle positions are given by
\begin{equation}
  \label{eq:4frm5}
  \begin{gathered}
  \mat{R}_{AD} = \sqrt{t_1} \ver{Z}~,
  \qquad
  \mat{R}_{CD} = \frac{\sqrt{t_1t_2-t_3^2}}{\sqrt{t_1}} \ver{Y} 
  + \frac{t_3}{\sqrt{t_1}} \ver{Z}~,\\
  \mat{R}_{BD} = \frac{t_6}{\sqrt{t_1t_2-t_3^2}} \ver{X}
  + \frac{t_1t_5-t_3t_4}{\sqrt{t_1}\sqrt{t_1t_2-t_3^2}}\ver{Y}
  +\frac{t_4}{\sqrt{t_1}} \ver{Z}~.
  \end{gathered}
\end{equation}
Particle position vectors $\mat{R}_{A,B,C,D}$ can of course be found
from (\ref{eq:4frm5}) together with the center-of-mass condition.  

The vibrational kinetic energy $\Kz$ in standard order  is determined
by the coefficients $g_{ab}$ and $g_b$ in (\ref{eq:kostan}).  Due to
the polynomial nature of $t_a$, the results for $g_b$ are remarkably
simple 
\begin{equation}
  \label{eq:4frm6}
  g_1=\frac{6}{\mu_{AD}}~,\quad
  g_2=\frac{6}{\mu_{CD}}~,\quad
  g_3=g_4=g_5=\frac{6}{\mu_{D}}~,\quad
  g_6=0~.
\end{equation}
The expressions for $g_{ab}$ are unavoidably more complicated, even
though their dependence on $\mat{R}_\alpha$ is polynomial.  Expressing
$g_{ab}$ in terms of internal coordinates $t_a$ we get, taking into
account its symmetry,
  \begin{gather}
    g_{11} = \frac{4}{\mu_{AD}} t_1~,~
    g_{12}= \frac{4}{\mu_{D}} t_3~,~
    g_{13}= \frac{2}{\mu_{AD}} t_3 + \frac{2}{\mu_{D}} t_1~,~
    g_{14}= \frac{2}{\mu_{AD}} t_4 + \frac{2}{\mu_{D}} t_1~,~
    g_{15}= \frac{2}{\mu_{D}} (t_3+t_4)~,\nonumber\\
    g_{16} = \frac{2}{\mu_{A}} t_6~,~
    g_{22} = \frac{4}{\mu_{CD}} t_2~,~
    g_{23} = \frac{2}{\mu_{CD}} t_3 + \frac{2}{\mu_D} t_2~,~    
    g_{24}= \frac{2}{\mu_{D}} (t_3+t_5)~,~
    g_{25}= \frac{2}{\mu_{CD}} t_5 + \frac{2}{\mu_{D}} t_2~,\nonumber\\
    g_{26} = \frac{2}{\mu_{C}} t_6~,~
    g_{33}= \frac{1}{\mu_{AD}} t_2 + \frac{1}{\mu_{CD}} t_1 +
    \frac{2}{\mu_D} t_3~,~
    g_{34}= \frac{1}{\mu_{AD}} t_5 + \frac{1}{\mu_D} (t_1+t_3+t_4)~,
    \nonumber\\ 
    g_{35}= \frac{1}{\mu_{CD}} t_4 + \frac{1}{\mu_D} (t_2+t_3+t_5)~,~
    g_{36}= 0~,~
    g_{44}= \frac{1}{\mu_{AD}} \mat{R}_{BD}^2 + \frac{1}{\mu_{BD}} t_1 +
    \frac{2}{\mu_D} t_4~,    \label{eq:4frm7}\\
    g_{45}= \frac{1}{\mu_{BD}} t_3 + \frac{1}{\mu_D}
    (\mat{R}_{BD}^2+t_4+t_5)~,~ g_{46}=0~,~
    g_{55}= \frac{1}{\mu_{BD}} t_2+ \frac{1}{\mu_{CD}} \mat{R}_{BD}^2 +
    \frac{2}{\mu_D} t_5~,\nonumber\\
    g_{56}=0~,~
    g_{66}= \frac{1}{\mu_{A}}(\mat{R}_{BD}^2 t_2-t_5^2)+
    \frac{1}{\mu_{B}} (t_2t_1-t_3^2) + \frac{1}{\mu_C} (\mat{R}_{BD}^2
    t_1 -t_4^2)~. \nonumber
  \end{gather}
Here we have left $\mat{R}_{BD}^2$ indicated for convenience, its
expression in terms of internal coordinates is given by
(\ref{eq:4frm5}). 

The vibrational-rotational coupling term $\Ko$ is given in
(\ref{eq:Kstand}) in terms of the coefficients $\D_{ak}$ and $\B_{k}$
(see (\ref{eq:Ktheta1}) and (\ref{eq:clfund})).  The expression for
$\D_{ak}$ can be written most compactly in terms of position vectors.
Its non-vanishing components are,
\begin{equation}
  \label{eq:4frm8}
  \begin{gathered}
  \D_{21} = -\frac{2}{\mu_D}\frac{R_{CDY}}{R_{ADZ}}~,~
  \D_{31} = -\frac{1}{\mu_{AD}}R_{CDY}~,~
  \D_{41} = -\frac{1}{\mu_{AD}}\frac{R_{BDY}}{R_{ADZ}}~,~
  \D_{42} = \frac{1}{\mu_{AD}}\frac{R_{BDX}}{R_{ADZ}}~,\\
  \D_{43} = \frac{1}{\mu_{AD}}\frac{R_{BDX}R_{CDZ}}{R_{ADZ}R_{CDY}}
  -\frac{1}{\mu_D} \frac{R_{BDX}}{R_{CDY}}~,~
  \D_{51} = -\frac{1}{\mu_D}\frac{R_{CDY}+R_{BDY}}{R_{ADZ}}~,~  
  \D_{52} = \frac{1}{\mu_D}\frac{R_{BDX}}{R_{ADZ}}~,\\
  \D_{53} = \frac{1}{\mu_{D}}\frac{R_{BDX}R_{CDZ}}{R_{ADZ}R_{CDY}}
  -\frac{1}{\mu_{CD}} \frac{R_{BDX}}{R_{CDY}}~,~
  \D_{61} = \frac{1}{\mu_A}\frac{R_{BDX}R_{CDZ}}{R_{ADZ}}~,\\
  \D_{62} = \frac{1}{\mu_A R_{ADZ}}(\mat{R}_{BD}\wedge
  \mat{R}_{CD})_X~,~   
  \D_{63} = \frac{
  R_{CDZ}}{R_{CDY}}\D_{62} + \frac{1}{\mu_C}
  \frac{R_{ADZ}R_{BDY}}{R_{CDY}}~. 
  \end{gathered}
\end{equation}
The coefficients $\B_{k}$, on the other hand, acquire a very simple
form because the first term in (\ref{eq:clfund}) vanishes, leaving
only the contribution from the second term,
\begin{equation}
  \label{eq:4frm9}
  \B_{1} = \frac{1}{R_{ADZ}^2 R_{CDY}}
  \left(-\frac{1}{\mu_{AD}}R_{CDZ} + \frac{1}{\mu_D} R_{ADZ}\right) =
  \frac{1}{\sqrt{t_1t_2-t_3^2}}
  \left(-\frac{1}{\mu_{AD}}\frac{t_3}{t_1} + 
  \frac{1}{\mu_D} \right)~, 
\end{equation}
and $\B_{2}=0=\B_{3}$.

Finally, the Jacobian $\widetilde{\J}$ in the inner
product (\ref{eq:inner2}) can be computed to give,
\begin{equation}
  \label{eq:4frm10}
  \frac{1}{\widetilde\J} = 4 (\mat{R}_{AD}\wedge\mat{R}_{CD})^2 + 4
  \frac{\mu_B}{M} (\mat{R}_{AD}\wedge\mat{R}_{CD})\cdot
  (\mat{R}_{AD}\wedge\mat{R}_{BD}+\mat{R}_{BD}\wedge\mat{R}_{CD} +
  \mat{R}_{CD}\wedge\mat{R}_{AD})~.
\end{equation}
$\widetilde\J$ is singular at $\mat{R}_{AD}\wedge\mat{R}_{CD} = 0$, as
expected from our choice of frame and internal coordinates.  This
singularity, together with the somewhat involved integration limits
resulting from (\ref{eq:4frm4}), make the expression for the inner
product computationally cumbersome.  For systems whose equilibrium
configuration is far from the singularity, however, the contribution
from that region should be strongly suppressed by the wave functions
in (\ref{eq:inner2}).

\section{Final remarks}
\label{sec:final}

In this paper we derived the body-frame Hamiltonian for a system of
$N$ particles in terms of frame conditions and internal coordinates.
Obtaining the Hamiltonian in terms of frame conditions instead of
Euler angles and the inertia tensor and their derivatives leads
arguably to computational simplifications.  All frames used in
applications are defined by polynomial conditions, usually of first or
second degree.  The coefficients $\N$, $\D$ and $\B$ in the kinetic
energy operator (\ref{eq:Kstand}), given by algebraic expressions in
terms of first derivatives of those frame conditions, can be
efficiently evaluated with symbolic computer algorithms or, depending
on $N$ and the internal coordinates $t_a$ being used, even by hand.
In particular, there is no need to invert the inertia tensor, or to
compute its determinant or that of the vibrational kinetic tensor
$g_{ab}$ (\ref{eq:kostan}).  Similarly, neither those determinants nor
derivatives of Euler angles are required for the computation of the
volume element in the quantum inner product as given by
(\ref{eq:detqr}) and (\ref{eq:detqr1}), and the derivatives of
internal coordinates involved in (\ref{eq:detqr1}) are evaluated only
at the frame manifold.  Those simplifications should be more apparent
the larger the value of $N$.
Furthermore, given a set of internal
coordinates $\{t_a\}$, it is straightforward to compute the
Hamiltonian in different frames by changing the conditions $\G_a$, as
illustrated in the examples of section \ref{sec:examp}.

The Hamiltonian is given in standard order in (\ref{eq:Kstand}) and in
the alternate forms (\ref{eq:covariant2}) and (\ref{eq:covariant3}).
Comparing those three forms leads to some useful relations among their
coefficients, in particular the expression (\ref{eq:A}) for the gauge
field $\mat{A}_a$ in terms of frame conditions.  One advantage of the
standard-order form (\ref{eq:Kstand}) is that it is known to be
equivalent to a path-integral formulation in phase space with
post-point discretization, whereas for the undefined orderings of
(\ref{eq:covariant2}) and (\ref{eq:covariant3}) the path-integral
equivalents are in principle not known and have to be constructed.  In
section \ref{sec:matrig}, in connection with the quantum inner product
in the body-frame, we discuss two equivalent representations for the
angular momentum operators and the kinetic energy. Namely, as
irreducible matrices acting on $\ell$-component wave functions,
(\ref{eq:matham}), and as differential operators acting on
rigid-rotator wave functions, (\ref{eq:rotham}).  The rigid-rotator
representation, which can be a convenient alternative to the matricial
one for some computations, appears naturally in the gauge-invariant
approach of~\cite{bou04}.

In section \ref{sec:preli} we discuss frame conditions from the point
of view of their admissibility and reparametrizations.  Not discussed
in this paper is the problem of frame singularities.  From
(\ref{eq:admi}) we see that the singular points on the frame manifold
are determined by the equations $\det(Q_{ai}(\{\mat{R}\})) = 0
=\G_a(\{\mat{R}\})$, which are polynomial in $\mat{R}$ for polynomial
$\G_a$.  Such algebraic formulation of the problem might be useful in
the study of frame singularities for larger values of $N$.

Applications of the approach presented here to the analyisis of
systems with $N>3$ are currently in progress, and will be discussed
elsewhere.

\appendix

\section{Remarks on the body-frame transformation}
\label{sec:app0}

\setcounter{equation}{0}
\renewcommand{\theequation}{\thesection.\arabic{equation}}

Equation (\ref{eq:Ufund}) for $\partial\mat{U}/\partial\mat{r}_\alpha$
gives a relation between the dependence of $\mat{U}(\{\theta_b\})$ on
$\mat{r}_\alpha$ and the frame conditions $\G_a$.  Notice that on the
r.h.s.\ of (\ref{eq:Ufund}) the dependences on $\{t_a\}$ and
$\{\theta_b\}$ are completely factorized, with $\F_{m\alpha n}$
depending only on $\{t_a\}$.  We can rewrite (\ref{eq:Ufund}) as,
\begin{equation}
  \label{eq:Ufund2}
  \F_{i\alpha j} = \frac{1}{2}\varepsilon_{ikl} \frac{\partial
  U_{km}}{\partial r_{\alpha n}} U_{lm} U_{jn}~.
\end{equation}
Differentiating both sides of (\ref{eq:Ufund2}) and using
(\ref{eq:euclid}) and (\ref{eq:32b}) we obtain,
\begin{equation}
  \label{eq:Ufund3}
  \frac{\partial\F_{i\alpha j}}{\partial R_{\alpha j}} = \frac{1}{2}
  \varepsilon_{ikl} \frac{\partial^2 U_{kn}}{\partial r_{\alpha m}
  \partial r_{\alpha m}} U_{ln}~.
\end{equation}
Notice that the l.h.s.\ of (\ref{eq:Ufund2}) and (\ref{eq:Ufund3})
depend only on $\{t_a\}$.  The l.h.s.\ of (\ref{eq:Ufund3}) appears in
the coefficient $\B_l$ defined by (\ref{eq:clfund}).

\section{Some useful identities}
\label{sec:appa}

In this appendix we gather some useful relations analogous to
(\ref{eq:lemma2-p15}) \cite{lit97}.  The last of these,
(\ref{eq:lemma4-p17}), is a closure relation which must hold on the
frame manifold (\ref{eq:frame}), except at singular points.
\begin{align}
  g^{-1}_{ab} &= h_{ab} - a_{ai} a_{bj} M^{-1}_{ij}
  \label{eq:lemma5-p18}\\ 
  \N_{ij} &= M_{ij} - \sum_{c,d=1}^{3N-6} h^{-1}_{cd} a_{ci}a_{dj}  
  \label{eq:lemma3-p16}\\
  \sum_{b=1}^{3N-6} g_{ab}a_{bi} &= M_{ij} \N^{-1}_{jk}
  \sum_{b=1}^{3N-6} h^{-1}_{ab} a_{bk} \label{eq:corollary1-p16.1}\\
  \N^{-1}_{ij} &= M^{-1}_{ij} + \sum_{a,b=1}^{3N-6} A_{ai}
  g_{ab} A_{bj}\label{eq:corollary2-p16.1}\\
  h^{-1}_{ab} &= g_{ab} -\sum_{c,d=1}^{3N-6} g_{ac} A_{ci}\N_{ij}
  A_{dj} g_{db} \label{eq:corollary3-p16.1}\\
  \frac{\det(\mat{M})}{\det(g_{ab})} &=
  \det(h_{ab})\det(\N)\label{eq:lemma6-p18} \\
  m_\gamma\delta_{\gamma\beta}\delta_{jk} &=
  \sum_{a,b=1}^3\R^{-2}_{ab} \frac{\partial\G_a}{\partial\R_{\gamma
  k}} \frac{\partial\G_b}{\partial\R_{\beta j}} + \sum_{c,d=1}^{3N-6}
  m_\gamma \frac{\partial R_{\gamma k}}{\partial t_c} h^{-1}_{cd}
  m_\beta \frac{\partial R_{\beta j}}{\partial t_d} + \delta_{jk}
  \frac{m_\gamma m_\beta}{M}\label{eq:lemma4-p17}
\end{align}

\begin{bibsection}[References]
  \begin{biblist}
\bib{bou04}{article}{  
author={Bouzas, A. O.},        
author={M\'endez Gamboa, J.},        
date={2004},          
journal={J. Phys. A},       
volume={37},        
pages={6773}}
%
\bib{han87}{article}{  
author={Handy, N. C.},        
date={1987},          
journal={Mol. Phys.},       
volume={61},        
pages={207}}
%
\bib{lit97}{article}{  
author={Littlejohn, R. G.},        
author={Reinsch, M.},        
date={1997},          
journal={Rev. Mod. Phys.},       
volume={69},        
pages={213}}
%
\bib{mer03}{article}{  
author={Meremianin,  A. V.},        
author={Briggs, J. S.},        
date={2003},          
journal={Phys. Rep.},       
volume={384},        
pages={121}}
%
\bib{eck35}{article}{  
author={Eckart, C.},        
date={1935},          
journal={Phys. Rev.},       
volume={47},        
pages={552}}
%
\bib{lou76}{article}{  
author={Louck, J. D.},        
author={Galbraith, H. W.},        
date={1976},          
journal={Rev. Mod. Phys.},       
volume={48},        
pages={69}}
%
\bib{men03}{article}{  
author={M\'endez Gamboa, J.},        
author={Bouzas, A. O.},        
date={2003},          
journal={J. Phys. A},       
volume={36},        
pages={7061}}
%
\bib{brn}{book}{   
author={Brink, D. M.},      
author={Satchler, G. R.},      
title={Angular momentum},       
date={1993},        
publisher={Oxford U. Press},   
place={New York},}      
%
\bib{wat68}{article}{  
author={Watson, J. K. G.},        
date={1968},          
journal={Mol. Phys.},       
volume={15},        
pages={479}}
%
\bib{car83}{article}{  
author={Carter, S.},        
author={Handy, N. C.},        
author={Sutcliffe, B. T.},        
date={1983},          
journal={Mol. Phys.},       
volume={49},        
pages={745}}
%
\end{biblist}
\end{bibsection}
\end{document}